\PassOptionsToPackage{hyphens}{url}
\documentclass[twocolumn]{aastex7}

\usepackage[colorinlistoftodos]{todonotes}

\usepackage{bm}
\usepackage{amsmath}
\usepackage{amssymb}
\usepackage{graphicx} 
\usepackage{float} 
\usepackage{subfigure}
\usepackage{enumerate}
\usepackage{booktabs}
\usepackage{listings}
\usepackage{makecell}
\usepackage{enumitem}
\usepackage{placeins}

\allowdisplaybreaks[2] 
\hypersetup{colorlinks=true, citecolor=blue, linkcolor=blue,
  urlcolor=magenta} \makeatletter
\newcommand{\rmnum}[1]{\romannumeral #1}
\newcommand{\Rmnum}[1]{\expandafter\@slowromancap\romannumeral
  #1@} \makeatother

\newcommand{\code}[1]{\lstinline{#1}}
\newcommand{\kratos}{\code{Kratos}}

\newcommand{\NetA}{Net-3Sp{}}       
\newcommand{\NetB}{Net-13Sp{}}      

\begin{document}

\title{Deep Neural Networks for Modeling Astrophysical Nuclear Reacting Flows}

\correspondingauthor{Yuxiao Yi, Lile Wang and Yao Zhou}
\email{yiyuxiao@sjtu.edu.cn, lilew@pku.edu.cn, yao.zhou@sjtu.edu.cn}
 
\author[0009-0006-4343-3465]{Xiaoyu Zhang}
\affiliation{School of Physics and Astronomy, Shanghai Jiao Tong University, Shanghai 200240, China}
\affiliation{Institute of Natural Sciences and MOE-LSC, Shanghai Jiao Tong University, Shanghai 200240, China}
\email{xiaoxiangyeyu@sjtu.edu.cn}

\author[0009-0002-6409-2760]{Yuxiao Yi}
\affiliation{School of Mathematical Sciences, Shanghai Jiao Tong University, Shanghai 200240, China}
\affiliation{Institute of Natural Sciences and MOE-LSC, Shanghai Jiao Tong University, Shanghai 200240, China}
\email{yiyuxiao@sjtu.edu.cn}

\author[0000-0002-6540-7042]{Lile Wang}
\affiliation{The Kavli Institute for Astronomy and Astrophysics, Peking University, Beijing 100871, China}
\affiliation{Department of Astronomy, School of Physics,
  Peking University, Beijing 100871, China} 
\email{lilew@pku.edu.cn}

 \author[0000-0003-0627-3520]{Zhi-Qin John Xu}
 \affiliation{School of Mathematical Sciences, Shanghai Jiao Tong University, Shanghai 200240, China}
\affiliation{Institute of Natural Sciences and MOE-LSC, Shanghai Jiao Tong University, Shanghai 200240, China}
\email{xuzhiqin@sjtu.edu.cn}

 \author[0000-0003-4639-9910]{Tianhan Zhang}
 \affiliation{School of Astronautics, Beihang University, Beijing 100191, China}
 \affiliation{Key Laboratory of Spacecraft Design Optimization and Dynamic Simulation Technologies, Ministry of Education, Beijing, 102206, China}
\email{thzhang@buaa.edu.cn}

\author[0000-0002-3616-2912]{Yao Zhou}
\affiliation{School of Physics and Astronomy, Shanghai Jiao Tong University, Shanghai 200240, China}
\affiliation{Institute of Natural Sciences and MOE-LSC, Shanghai Jiao Tong University, Shanghai 200240, China}
\email{yao.zhou@sjtu.edu.cn}



\begin{abstract}
In astrophysical simulations, nuclear reacting flows pose computational challenges due to the stiffness of reaction networks. We introduce neural network-based surrogate models using the DeePODE framework to enhance simulation efficiency while maintaining accuracy and robustness. Our method replaces conventional stiff ODE solvers with deep learning models trained through evolutionary Monte Carlo sampling from zero-dimensional simulation data, ensuring generalization across varied thermonuclear and hydrodynamic conditions. Tested on 3-species and 13-species reaction networks, the models achieve $\lesssim 1\%$ accuracy relative to semi-implicit numerical solutions and deliver a $\sim 2.6\times$ speedup on CPUs. A temperature-thresholded deployment strategy ensures stability in extreme conditions, sustaining neural network utilization above 75\% in multi-dimensional simulations. These data-driven surrogates effectively mitigate stiffness constraints, offering a scalable approach for high-fidelity modeling of astrophysical nuclear reacting flows.
\end{abstract}


 
\section{Introduction}
\label{sec:intro}

Nuclear reacting hydrodynamic flows play a crucial role in astrophysical phenomena such as stellar evolution, supernova explosions, and extreme processes near compact objects. Numerical simulations of these flows require coupling fluid dynamics with nuclear reaction networks, including the advection of nuclear species and energy feedback mechanisms. Recent advancements in computational power and multi-physics algorithms have
enabled increasingly sophisticated modeling of these
systems. Current research focuses on integrating multi-scale
physics to resolve challenges such as turbulent combustion,
neutrino-matter interactions, and high-density
equation-of-state dependencies. Numerical systems for
nuclear reacting flows have been instrumental in studying
double detonation scenarios \citep{townsley_double_2019,
  gronow_double_2021, rivas_impact_2022,
  ghosh_confronting_2022}, deflagration-to-detonation
transitions \citep{gamezo_deflagrations_2004,
  gamezo_threedimensional_2005}, white dwarf collisions
\citep{kushnir_head-collisions_2013,
  garcia-senz_high-resolution_2013}, and X-ray bursts on
neutron stars \citep{Zingale_2023,
  johnson2024simulatinglateralhheflame,
  guichandut_hydrodynamical_2024}.

The numerical treatment of nuclear reacting flows remains
challenging due to extreme stiffness arising from the timescale
separations spanning orders of magnitude between nuclear
reactions. To ensure stability and accuracy, implicit or
semi-implicit ordinary differential equation (ODE) solvers
with adaptive step-size control are typically
employed. These methods necessitate computationally
expensive evaluations of reaction rate vectors, Jacobian
matrices, and matrix inversions. Established simulation
tools such as \code{Castro} \citep{zingale_improved_2019}, \code{MAESTROeX} \citep{2019ApJ887212F}, and
\code{FLASH} \citep{Fryxell_2000} have adopted this
paradigm, enabling critical insights into astrophysical
systems over decades of development. However, prohibitive
computational costs continue to limit explorations requiring
finer reaction-network details or higher spatiotemporal
resolutions, underscoring the need for optimized algorithms
and high-performance computing strategies.

Recent advances in deep learning (DL) techniques have
demonstrated potential to enhance nuclear reacting flow
simulations by replacing traditional semi-implicit solvers
with data-driven surrogate models based on deep neural
networks (DNNs). This approach has gained traction in
combustion chemistry simulations \citep{IHME2022101010}, 
where \citet{zhang2022CF,WANG2025114105} pioneered DNN-based
surrogate models capable of advancing hydrodynamic and
thermochemical time steps while circumventing stiffness
inherent to chemical kinetic networks. These models have
been validated across diverse reacting flow scenarios,
achieving order-of-magnitude speedups compared to
conventional solvers without compromising accuracy. Building
on this, \citet{deepode2025} introduced \code{DeePODE}, a
novel framework combining evolutionary Monte Carlo sampling
with neural networks to tackle multi-timescale systems such
as electrochemical battery thermal runaway and turbulent jet flame. 
At its core, the method leverages short-term ODE integration to embed multiscale characteristics into the sampled data, facilitating the efficient construction of surrogate models for high-dimensional stiff systems. Given the mathematical
equivalence between nuclear reaction networks and chemical
kinetics, this paradigm holds significant promise for
accelerating simulations in nuclear astrophysics. 

Despite these developments, the application of DL-based
surrogate models to nuclear astrophysical systems remains
underexplored. \citet{fan_neural_2022} demonstrated a
proof-of-concept by training a DNN to replace nuclear
reaction solvers in flame simulations at
$\sim 10^{-1}~{\rm cm}$ scales involving 3 nuclear species.
\cite{smith_machine_2024} discussed the importance of employing neural networks to solve nuclear burning in multidimensional nuclear reacting flow simulations, and outlined the various challenges that currently exist in this approach. Subsequently, \cite{grichener_nuclear_2025} employed neural networks to replace the reaction network solver for approximately 100 isotopes in a zero-dimensional case, achieving errors of no more than 1\% per time step.
Although their work highlighted feasibility and
computational gains, significant challenges persist in
adopting such methods to high-dimensional reaction
networks and extreme parameter regimes typical of
astrophysical environments. In this study, we build upon the sampling strategy introduced in \code{DeePODE} framework \citep{deepode2025} and adopt its neural network architecture as the foundation for our surrogate modeling approach to address
these challenges, structuring our methodology into three
components: (1) Monte Carlo sampling of nuclear reaction
data, (2) DNN training, and (3) validation within reacting
flow simulations. Unlike prior studies, our sampling
strategy is decoupled from specific flame configurations,
enabling a single trained model to generalize across broad
regions of parameter space without fine-tuning—a critical
requirement for astrophysical applications.

This paper is structured as follows. The equations to
be solved are described in \S\ref{sec:eqn}. The neural network architecture, sampling methodology, and training procedures are presented in \S\ref{sec:dnn}. Building on this methodological foundation, \S
\ref{sec:results} conducts systematic comparisons of the calibration accuracy in the zero-, one-, and two-dimensional simulation frameworks and performs acceleration performance. Finally, \S\ref{sec:summary} concludes with a summary of key findings and outlines potential directions for future research.

\section{Governing equations of nuclear reacting flows}
\label{sec:eqn}

Reacting flow simulations are carried out based on
hydrodynamic calculations carrying reacting species,
\begin{align}
  &\frac{\partial (\rho X_{k})}{\partial t} + \nabla \cdot
    (\rho X_{k} \bm{v}) = \rho
    \dot{\omega}_{{k}}\ , \label{eq:mass} \\ 
  &\frac{\partial(\rho\bm{v})}{\partial t} + \nabla
    \cdot(\rho\bm{v}\bm{v}+p\bm{I}) = 0\ ,
    \label{eq:momentum} \\ 
  &\frac{\partial(\rho\epsilon)}{\partial t} + \nabla
    \cdot[\bm{v}(\rho\epsilon+p)] =  \rho \dot
    S\ . \label{eq:energy} 
\end{align}
Here $\rho$, $p$, $\bm{v}$ are the mass density, the gas
pressure, and the velocity, respectively, $X_{k}$ is the
mass fraction of species $k$ ($\sum_k X_k = 1$), whose
associated production rate is $\dot{\omega}_{{k}}$, and
$\bm{I}$ is the identity tensor. In Eq.(\ref{eq:energy}),
$\epsilon$ denotes the total specific energy and $\dot S$ is
the rate of generation of nuclear energy (possibly including thermal neutrino losses) per unit
mass. In this paper, these equations are solved using the
\kratos{} framework (see \citealt{wang2025}, and also
\S\ref{sec:results}), on which the nuclear reaction
calculations are carried out.

Nuclear reactions and hydrodynamics can be coupled via the
operator splitting scheme (such as Strang splitting
\citep{Strang1968}), by which they can be computed
separately. The governing ODEs for nuclear reactions are:
\begin{align}
& \frac{\mathrm{d} \bm{X} }{\mathrm{d} t} =
  \bm{\dot{\omega}} (T, \rho, \bm{X})\ \label{eq_mass_fraction},\\ 
&\frac{\mathrm{d} (\rho\epsilon)}{\mathrm{d} t}  =  \rho
  \dot S\ \label{eq_energy},
\end{align} 
where $\bm{X}$ denotes the vector of mass fractions of species, and the production rates $\bm{\dot{\omega}}$ are generally associated with the temperature $T$ and density $ \rho$. The temperature is updated using the equation of state (EOS). Although complex equations of state are commonly used in astrophysical simulations, this study aims primarily to evaluate the performance of the neural network. Therefore, a simplified gamma-law EOS is adopted for all stages, including data sampling, training, and calibration. For the gamma-law EOS, 
\begin{equation}
    \begin{aligned}
    p = \rho \epsilon_{int} (\gamma-1)\ ,
    \end{aligned}
\end{equation}

we adopt the ideal gas law with an adiabatic coefficient of $\gamma=5/3$, and specific internal energy $\epsilon_{int}$ can be calculated by subtracting the specific kinetic energy from the total specific energy $\epsilon$. 


In standard computational workflows, the integral time step for advancing nuclear reaction networks is significantly smaller than that for hydrodynamic processes (namely $\Delta t_{\text{react}} \ll \Delta t_{\text{flow} }$), leading to substantial computational overhead. Therefore, this study focuses on developing surrogate models to circumvent stiffness constraints, thereby seamlessly replacing the time-consuming direct integration (DI).

\begin{figure*}[htbp]
\centering  
\subfigure[]{
\label{Fig.sp3}
\includegraphics[width=0.45\textwidth]{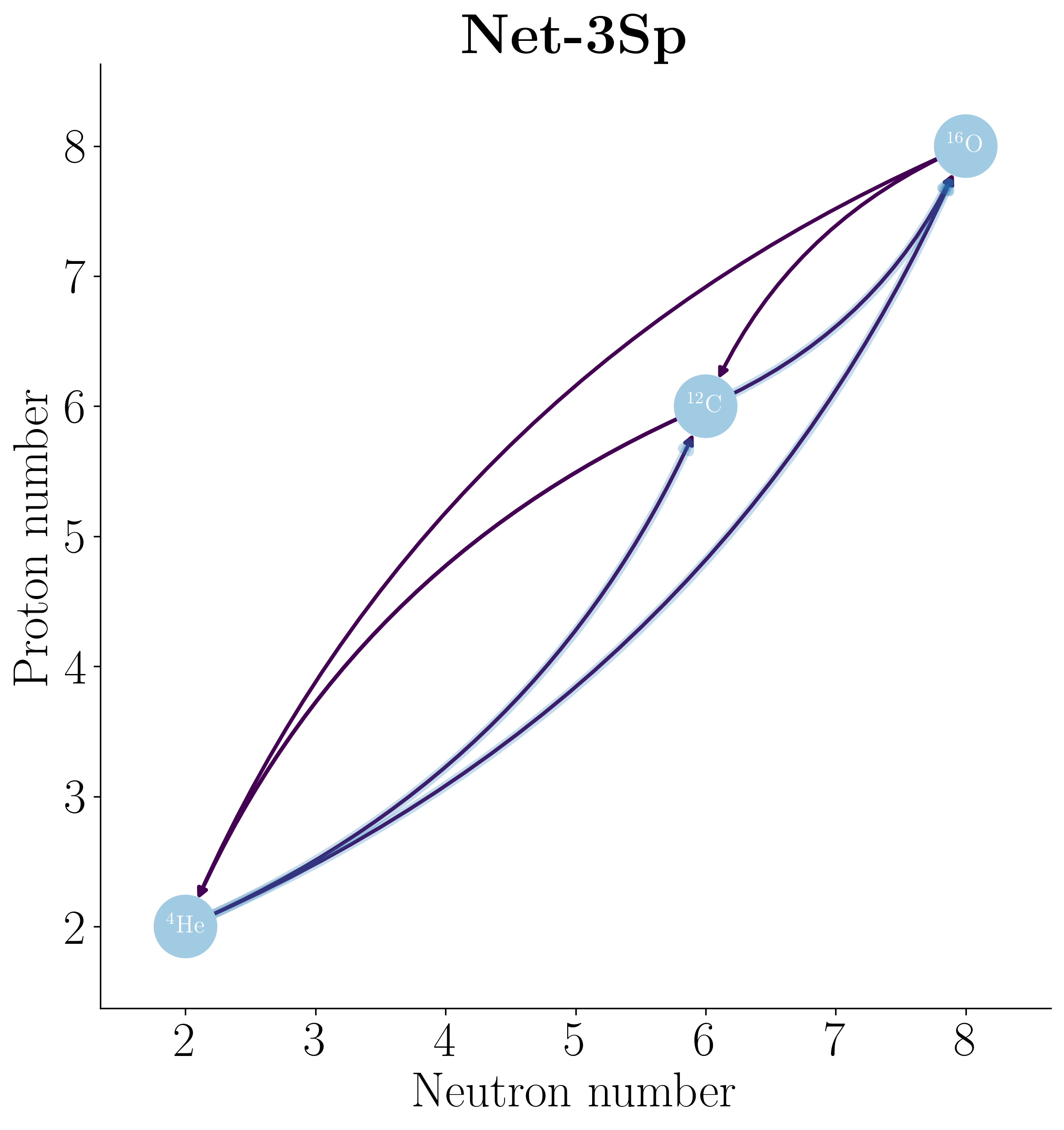}}
\subfigure[]{
\label{Fig.sp13}
\includegraphics[width=0.45\textwidth]{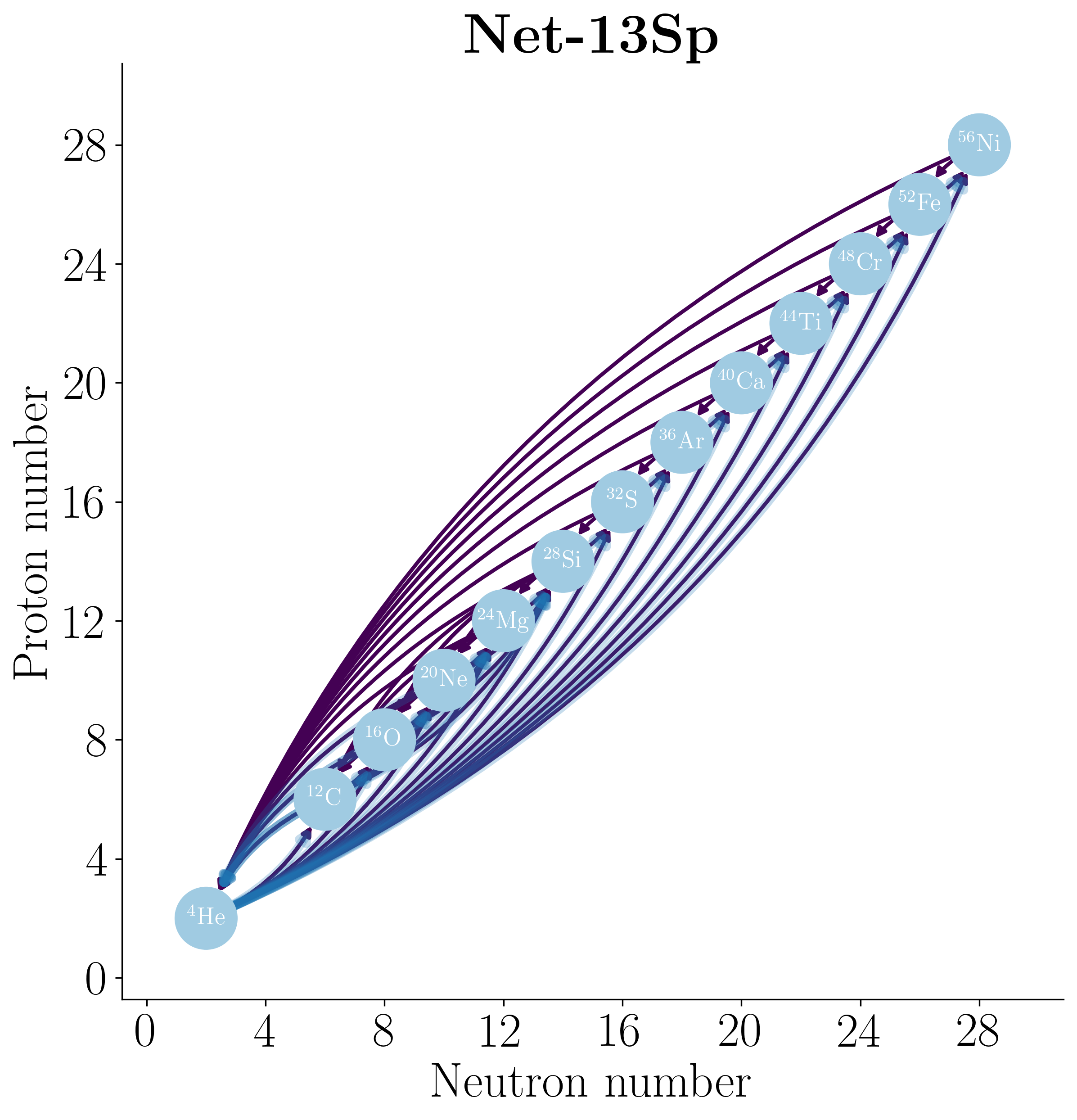}}
\caption{The reaction diagrams of two nuclear reaction networks, both generated by pynucastro, with the arrows for exothermic reactions marked in blue. (a) Reaction schematic of the \NetA, includes two forward reactions and two reverse reactions. (b) Reaction schematic of the \NetB, includes sixteen forward reactions and sixteen reverse reactions.}
\label{Fig.networks}
\end{figure*}

In our work, we select two distinct nuclear reaction networks. 
The small-size network, termed \NetA, is a simplified test network that includes 3 isotopes ($^{4}$He, $^{12}$C, $^{16}$O) and 4 reactions, as illustrated in Figure \ref{Fig.sp3}. 
The moderate-size network, \NetB, is a more comprehensive model designed for simulations of white dwarf combustion \citep{rivas_impact_2022}. 
It encompasses thirteen isotopes ($^{4}$He,$^{12}$C,$^{16}$O,$^{20}$Ne,$^{24}$Mg,$^{28}$Si,$^{32}$S,$^{36}$Ar,$^{40}$Ca,$^{44}$Ti,\allowbreak $^{48}$Cr,$^{52}$Fe,$^{56}$Ni) and incorporates thirty-two nuclear reactions, as depicted in Figure \ref{Fig.sp13}. This detailed network is particularly suited for modeling the complex nucleosynthesis and energy generation processes in astrophysical scenarios. 
In contrast to the method adopted in the \cite{timmes_integration_1999} , the $\rm{(\alpha,p)(p,\gamma)}$ links are not approximately incorporated into the reaction network. In addition, the adopted reaction networks do not account for any reactions that involve the generation or annihilation of neutrinos, \added{and the thermal neutrino loss processes such as the pair neutrino process, photoneutrino process, plasma neutrino process, and bremsstrahlung neutrino process \citep{itoh_neutrino_1996} are also not included here}. 
The coefficients for these reaction networks are generated using the pynucastro \citep{smith_pynucastro_2023, pynucastro, the_pynucastro_development_2024_13899727}, which constructs reaction networks and provides essential nuclear reaction parameters such as reaction rates.

\section{Deep Neural Networks}
\label{sec:dnn}

\subsection{Architecture}
\label{sec:dnn-arch}
The temporal evolution of nuclear reaction under a desired time step $\Delta t$ can be written as:
\begin{equation}
    \begin{aligned}
    &\left[ \begin{array}{c}
    \bm{X}(t+\Delta t) \\ \epsilon(t+\Delta t) \end{array} \right] := \mathcal{F}(T(\epsilon(t),\rho,\bm{X}(t)), \rho, \bm{X}(t)) \\
    &= \left[ \begin{array}{c}
    \bm{X}(t) + \int_{t}^{t+\Delta t }\bm{\dot{\omega}}\mathrm{d} t \\
    \epsilon(t) + \int_{t}^{t+\Delta t }\dot{S}\mathrm{d} t
    \end{array} \right] \ ,
    \end{aligned}
\end{equation}
where $\mathcal{F}$ is a non-linear propagation operator, this operator simultaneously evolves the specific internal energy and the mass fractions. While specific energy and temperature can be inter-converted at times $t$ and  $t +\Delta t$ through the EOS, and they are linked by the specific heat capacity during the evolution process, \added{which is recomputed at each step to account for changes in physical conditions during the reaction.} We propose a deep neural network (DNN) model $\mathcal{F}_{\theta}$ to learn the mapping $\mathcal{F}$, and the $\theta$ denotes learnable parameters. In this work, for each neural network, the timestep $\Delta t$ is fixed and not treated as an explicit input. \added{The DNN model takes $[T, \rho, \bm{X}]$ as input and then predicts the temperature and the mass fraction of the species.} \added{Since the ODE integration depends on the EOS, any change in the EOS requires resampling the training data and retraining the neural network, and information from the EOS has been naturally absorbed by the neural network during training.}   

For the neural network structure, the feed-forward network (FFN) structure is considered in this study. Note that we employ an individual FFN for each nuclear reaction network. Each FFN comprises three hidden layers with 1,600, 800, and 400 neurons, respectively. The activation functions of hidden layers are  Gaussian Error Linear Unit (GELU).

As mentioned above, the input state vector is composed of temperature, density and mass fraction of species, denoted as $\bm{x}(t) = [T(t), \rho(t), \bm{X}(t)]$ with input size being $N_s+2$ where $N_s$ is the species size of reaction network. The label/target state of the DNN is $\bm{u}(t) = [ \bm{X}(t+\Delta t) - \bm{X}(t) ]$ with size = $N_s$, representing the change of mass fraction after a constant time step $\Delta t$. The training dataset is in the form of $\mathcal{D}_{\text{train}} = \{ \bm{x}_{i}, \bm{u}_{i} \}_{i=1}^{N} $. 

During the nuclear reaction processes, the mass fractions of
species exhibit typical multi-scale features. Across
distinct species or different evolutionary time, the
characteristic scales span multiple orders of magnitude,
ranging from $\mathcal{O}(10^{-25})$ to
$\mathcal{O}(10^{-1})$. Therefore, to ensure the stability
of DNN training, appropriate pre-processing of raw data is
required to transform the mass fraction to near
$\mathcal{O}(1)$ before directly feeding the dataset into
DNNs. Logarithm transformation seems natural, 
but fails to handle the scenarios when the concentrations
approach 0. Although we can add an offset $\varepsilon$
to the function to avoid discontinuities at zero, another
defect arises: sensitivity to perturbations. For instance,
assuming $\varepsilon = 10^{-25}$ and we define the function
$f(x) = \log_{10}(x+\varepsilon)$. If the input $x=10^{-20}$
is slightly perturbed to $\tilde{x}=10^{-30}$, this
operation results in a change $\Delta f\approx 5$. Such
machine-precision perturbation leads to significant
fluctuation, indicating that the logarithmic transformation
is ill-suited in our work.

To solve the prospective issue of multiple magnitude scales in predicting the reaction network, the Box-Cox
transformation (BCT) is introduced. It is a statistical method
originally proposed by \citet{Box1964} to stabilize variance
of data, and the function is defined as
\begin{align}
    B(x) = 
\begin{cases}
\displaystyle \frac{x^\lambda - 1}{\lambda} & \text{if } \lambda \neq 0\ , \\
\ln(x) & \text{if } \lambda = 0\ ,
\end{cases}
\end{align}
where $x$ is the input variable and $\lambda$ is a
hyper-parameter. It is widely used in regression analysis
and was first employed to deal with combustion chemistry
concentrations in the previous literature
\citep{zhang_dlode_2021}. BCT monotonically projects the mass
fractions within $[0,1]$ to $[-1/\lambda, 0]$ if
$\lambda >0$, and converts the low-order
concentrations $10^{-k}$ to $\mathcal{O}(1)$ magnitude,
avoiding the singularity and perturbation sensitivity when
approaching zero compared with the logarithm function. Hence, we
adopt BCT as a data pre-processing method with
$\lambda = 0.1$. In this study, we only apply BCT to species
mass fractions due to their inherent multi-scale
distribution. For other physical quantities such as
temperature and density, BCT pre-processing is deemed
unnecessary. Following the BCT transformation of mass
fractions, we implement Z-score normalization (i.e., subtract
the mean and divide by the standard deviation) across all
input and label the dataset to enhance the convergence speed of
DNNs.

\subsection{Data Generation and Sampling}
\label{sec:dnn-sampling}

In general, the performance of DNN training and
  prediction depends largely on the ``data quality'', which
  typically refers to the ability of training data to represent the situations in actual applications.  One
  common practice in improving data quality is to directly
  sample the data of reactions from reacting hydrodynamic
  simulations. However, for complex flame simulations, the
computational cost becomes prohibitive, especially in
three-dimensional cases involving
multiple components. Apart from this, the training dataset originates
from specific flame configurations, and the corresponding DNN
models could only be deployed into similar simulation
settings, thereby limiting the applicability. This is
because the simulation data represents low-dimensional
manifolds in the whole thermo-concentration space. The DNNs
are capable of learning the features within this manifold
distribution, while they perform poorly on the test samples
deviating from it. Such a phenomenon has been observed in
the previous work of \citet{fan_neural_2022}. To address these challenges, we need a sampling method that enhances model generalization capability, enables effective deployment of a single trained model across diverse testing scenarios, and maintains high prediction accuracy.

Compared to chemical reactions, nuclear reactions exhibit
significantly greater stiffness and heightened temperature
sensitivity. This characteristic not only necessitates the
collection of more extensive data from stiff regions, but
also results in substantially smaller time scales for
intense reaction dynamics, thereby compounding data
acquisition challenges. To overcome these issues, we extend
the Evolutionary Monte Carlo Sampling (EMCS)
\citep{deepode2025} for collecting samples from the broad phase
space and enhance data acquisition in highly rigid
regions. This method is performed independently of
hydrodynamics simulations, relying solely on the nuclear ODE
solver. The sampling procedure can be easily implemented in the code. The method comprises three main steps:
\begin{enumerate}
\item\label{item_U1} \textit{Range estimation}: Initially, we estimate the
  sampling ranges for temperature $T$, density $\rho$, and
  species mass fractions $X_k$ based on white dwarf
  deflagration scenarios.
    \item\label{item_U2} \textit{Monte Carlo sampling}: We conduct Monte
      Carlo sampling for $T$, $\rho$ and $X_k$ within the
      specified value ranges. This means that $\lg T$, $\lg\rho$ and $\lg X_k$ are randomly sampled on a logarithmic scale given by item \ref{item_U1} and an exponential operation is subsequently applied to obtain $T$, $\rho$ and $X_k$. Then the mass fractions are normalized, resulting in
      the formation of the state vector
      $\bm{x}_0(t) = [T, \rho, \bm{X}]$ and the variables are relatively evenly distributed across multiple scales.
    \item\label{item_U3} \textit{Data augmentation}: Using each state
      vector $\bm{x}_0(t)$ as an initial value for ODEs, we
      evolve the system along the ODE trajectory and collect
      data on it at the time points
      $[\tau_1, \tau_2, \cdots, \tau_{n}]$, namely
      $\bm{x}_i := \bm{x}_{i-1}(t+\tau_{i}), ~i=1,\cdots,
      n$. During the ODE integration, the EOS connects temperature and internal energy through heat capacities. After each calculation step, temperature and energy are converted back and forth using the EOS, ensuring consistency throughout the integration process. To better resolve stiff regions, time points are
      concentrated on the initial reaction stages where
      reaction stiffness typically peaks. Subsequently, all
      states $\bm{x}_i$ are consolidated to generate the
      corresponding labeled dataset.
\end{enumerate}

Specifically, we configure a
one-dimensional pipe with 10000 mesh cells and disable
the flow convection module to isolate nuclear reactions in Castro \citep{zingale_improved_2019}. And the initial value in each cell is generated by MC method described in item \ref{item_U2}. Nuclear reaction evolution is subsequently carried out in the one-dimensional pipe without convection, using the VODE solver for ODE integration. At the time point $\tau_{i}$ described in item \ref{item_U3}, the state of the entire pipe is extracted and independently evolved over a time interval $\Delta t$. For each time point, the values of [$T,\rho,\bm{X}$] in every mesh cell before and after evolution over $\Delta t$ form a training pair. This process yields 10,000 data samples per time point.

The value ranges for initial $[T, \rho, \bm{X}]$ used in item \ref{item_U2} and label time $\Delta$$t$ used in item \ref{item_U3}  for two networks are shown in Table \ref{table:sampling_range}. This process runs 500 times to generate a sufficient dataset. For each
network, this process yields nearly 100 million individual
data points. The data are extracted from Castro output files
by using yt interface \citep{turk_yt_2011}. To ensure the
dataset is suitable for training purposes, a filtering
process is applied based on the temperature gradient. The
rationale behind this filtering is to prioritize data with
larger temperature gradients, as these regions are often
more challenging for modeling nuclear reactions.

\begin{deluxetable}{ccc}
\label{table:sampling_range}
\tabletypesize{\small} 
\tablewidth{0.9\textwidth} 
\tablecaption{Variable ranges and label time for two networks}
\tablehead{
  \colhead{Variables} & \colhead{\NetA} & \colhead{\NetB} 
}
\startdata
$T/\rm{(10^{9}~K)}$ & [0.2,8] & [1,8] \\
$\rho/\rm{(10^{6}~g~cm^{-3})}$  & [2.5,100]& [20,200] \\
$X_{k}$  & [$10^{-6}$,1] & [$10^{-10}$,1]\\
$\Delta$$t/\rm{s}$ &$10^{-6}$&$10^{-7}$\\
\enddata
\tablecomments{See also \S \ref{sec:dnn-sampling}.}
\end{deluxetable}

\begin{figure}[htbp]
\centering  
\subfigure[]{
\label{Fig.phase_sp3}
\includegraphics[width=0.45\textwidth]{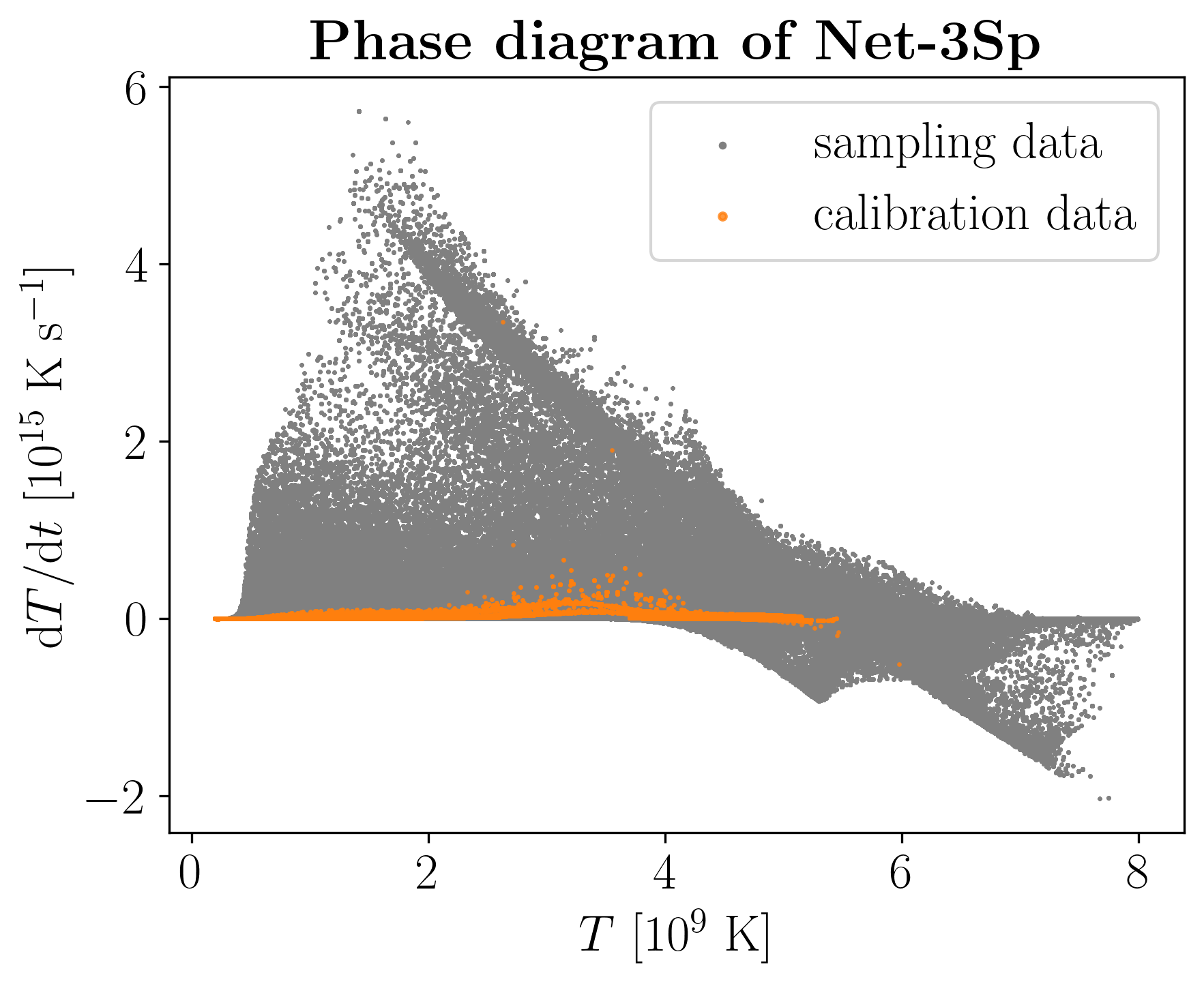}}
\subfigure[]{
\label{Fig.phase_sp13}
\includegraphics[width=0.45\textwidth]{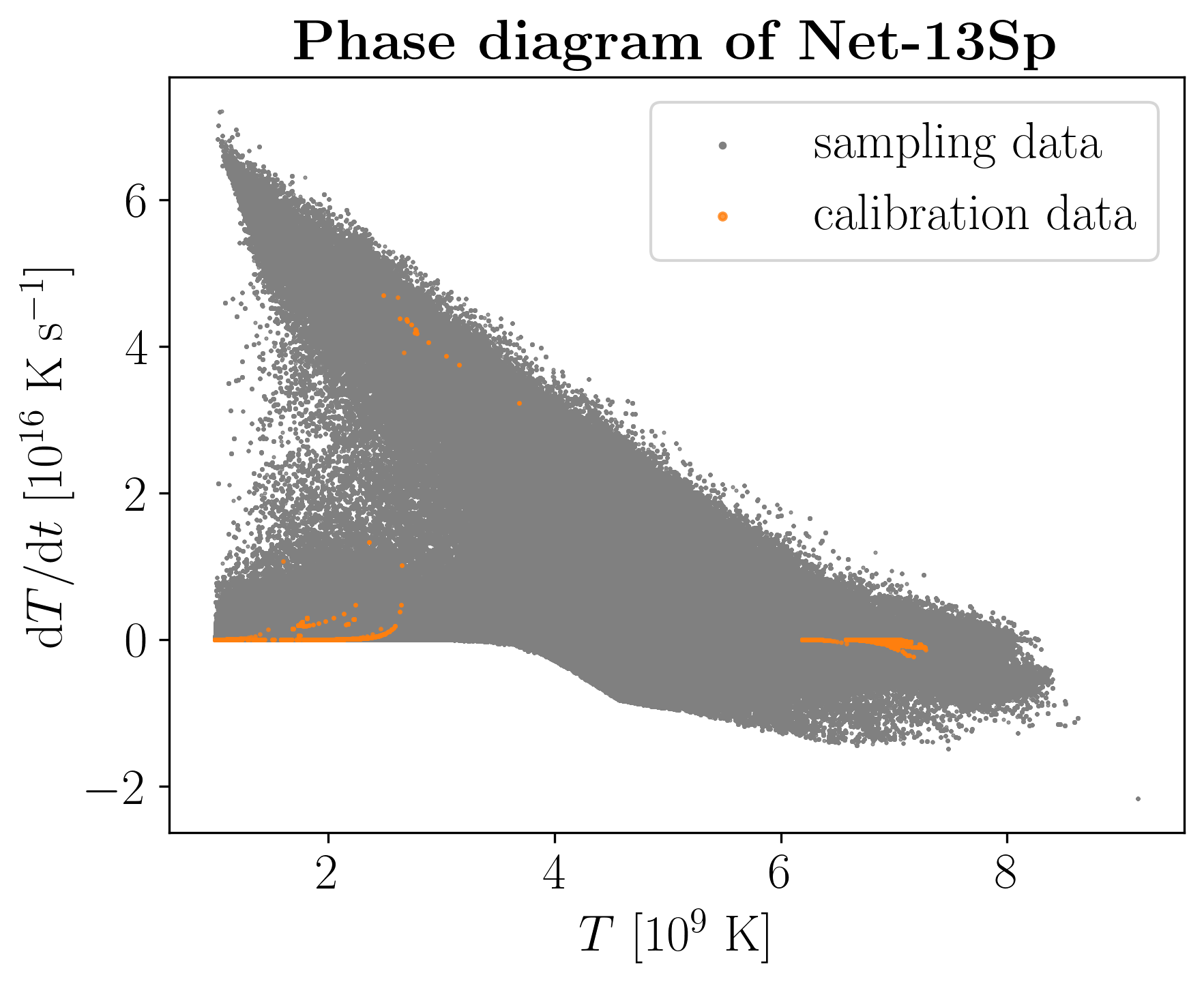}}
\caption{The phase diagrams of two nuclear reaction networks illustrate the temperature gradient (ordinate) against the temperature (abscissa). Gray points represent sampling data, while orange points are part of the calibrations data in \S\ref{sec:results}. (a) Phase diagram of the \NetA. (b) Phase diagram of the \NetB.}
\label{Fig.phase}
\end{figure}

To preserve a greater number of samples from stiff regions, we implement temperature gradient-based filtering during the selection process. For \NetA, 75\% of data points with temperature gradients exceeding $10^{13}~\mathrm{K~s^{-1}}$ are retained, whereas only 8\% retention occurs for gradients below this threshold. In contrast, \NetB{} demonstrate complete retention (100\%) for temperature gradients surpassing $8\times10^{13}~\mathrm{K~s^{-1}}$, with a sharp decline to 10\% retention rate for sub-critical gradient conditions. Finally, we obtain approximately 17 million data points for \NetA{} (see Figure \ref{Fig.phase_sp3}) and 28 million data points for \NetB{} (see Figure \ref{Fig.phase_sp13}), all filtered data obtained in this step are utilized for training the neural network. As shown in the figure, the sampling range essentially encompasses the data coming from the calibration simulations in \S\ref{sec:results}.


\subsection{Training}
\label{sec:dnn-training}
In order to perform training, the data is split into two subsets: 90\% for the training set (approximately 15.3 million data samples for \NetA{} and 25.2 million samples for \NetB) and 10\% for the validation set (approximately 1.7 million data samples for \NetA{} and 2.8 million samples for \NetB). The validation set is exclusively used to evaluate the model's performance and does not participate in the training process.
 
 The loss function is set as mean absolute error (i.e. $\text{L}_1$ loss), defined as 
\begin{align}
    L_{\text{data}} = \frac{1}{N}\sum_{i=1}^{N}\left\Vert \bm{u}_{i} - \tilde{\bm{u}}_{i} \right\Vert_{L_1}\ ,
\end{align}
where $ \tilde{\bm{u}}_{i}$ is the prediction with the input being $\bm{x}_i$ and $N$ is the training data size. Otherwise, the predicted mass fractions are expected to satisfy the physical constraint: $\sum_{k} \tilde{X}_k = 1$, and hence the corresponding loss is
\begin{align}
     L_{\text{sum}} = \frac{1}{N}\sum_{i=1}^{N}\left\Vert \sum_k^{N_s}\tilde{\bm{X}}_k^{i} - 1 \right\Vert_{L_1}\ .
\end{align}
The total loss for consideration will be 
\begin{align}
    L_{\text{all}} =   L_{\text{data}} +   L_{\text{sum}}
\end{align}
and the Adam algorithm is utilized for parameter optimization during the training procedure. The initial learning rate and batch size are $10^{-4}$ and 1024, respectively. After training for 2500 epochs, the learning rate is decayed by a factor of 10, and the batch size is increased by 128 times, continuing until reaching 5000 epochs. This two-stage training strategy is designed to enhance the generalization ability of the DNN in the first stage and accelerate convergence in the second stage. In the actual training process, we use PyTorch \citep{NEURIPS2019_9015}.

\begin{figure}[htbp]
\centering  
\includegraphics[width=0.45\textwidth]{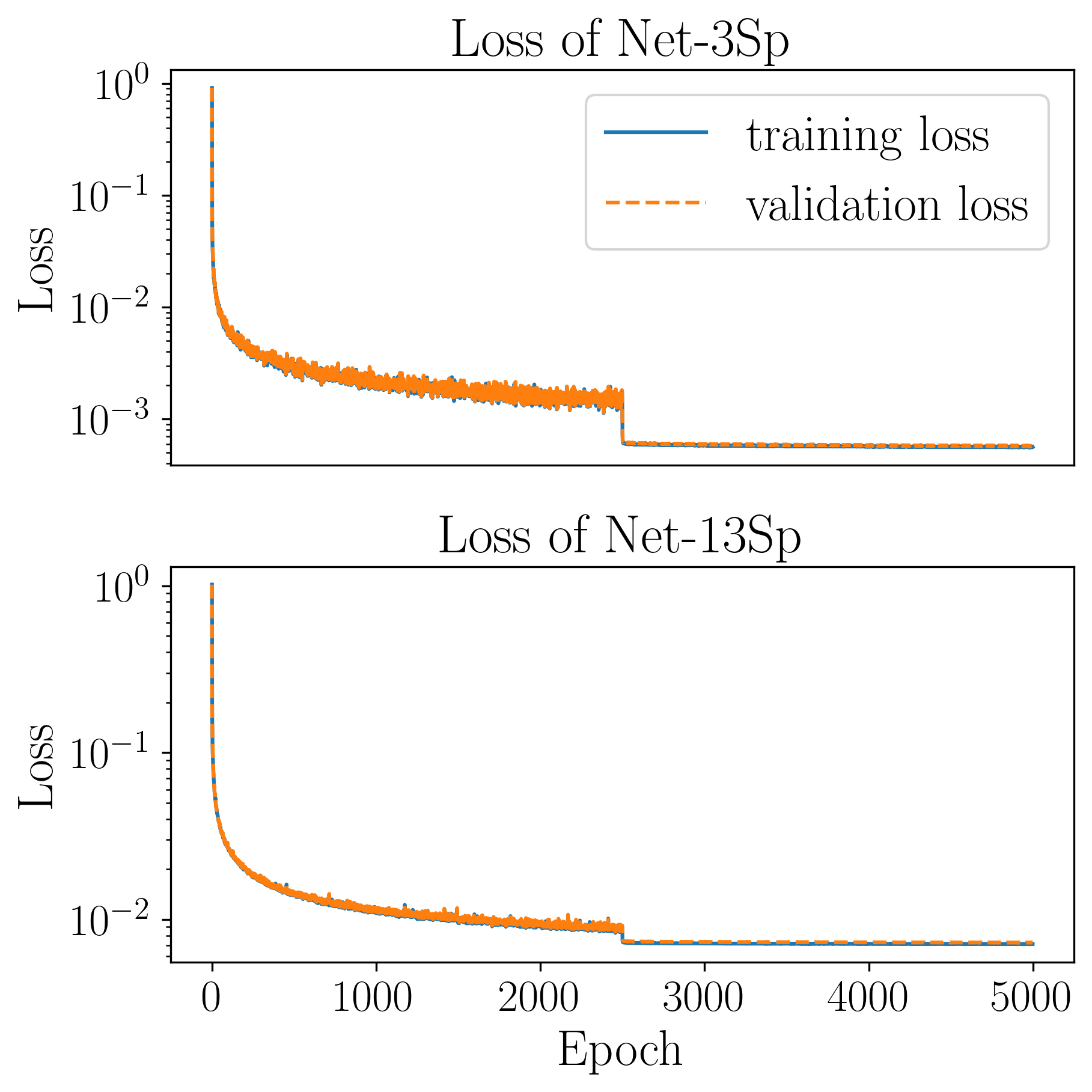}
\caption{L1 loss of two nuclear reaction networks, each training session consisted of 5000 epochs. The top panel is the loss of \NetA. The bottom panel is the loss of \NetB.}
\label{Fig.loss}
\end{figure}

As a result, in Figure
(\ref{Fig.loss}), after 2500 epochs, both the training loss
and the validation loss exhibits gradual
convergence to less than $10^{-3}$ for \NetA{} and less than $10^{-2}$ for \NetB{}. And the close alignment between training
and validation losses indicates the sufficiency of our
dataset for training. Finally, trained neural networks are
selected based on the number of epochs that yield optimal
performance: 2600 epochs for \NetA{} and 4000 epochs for
\NetB.

\section{Calibration results}
\label{sec:results}

To verify the accuracy of the neural network, we perform
calibrations in zero-dimensional, one-dimensional, and
two-dimensional scenarios. For one-dimensional and two-dimensional calibration, we
integrate the neural network into \kratos{} \citep{wang2025}
using PyTorch \citep{NEURIPS2019_9015}, then simulate and
calibrate using both the neural network and the direction
integration (solving Equation \ref{eq_mass_fraction} and \ref{eq_energy}), during each step of the direct integration, the energy evolves and is coupled to the temperature via the specific heat capacity given by equation of state (gamma-law in this work),  \kratos{} is a newly developed astrophysical
fluid dynamics code optimized for heterogeneous computing
architectures. \kratos{} provides a scalable,
high-performance computational framework for complex
simulations, emphasizing graphics processing unit (GPU)
efficiency and modular extensibility. Alternatively, with
the help of the
HIP-CPU\footnote{\url{https://github.com/ROCm/HIP-CPU}}
library, \kratos{} can be easily deployed on CPU
architectures. To resolve compressible reacting flows,
\kratos{} implements a finite-volume Godunov method, and
involves its own nuclear reaction models to calibrate the
reactions against. The numerical scheme combines Piecewise
Linear Method (PLM) spatial reconstruction with HLLC Riemann
solvers, providing second-order accuracy while preserving
positivity for multi-species systems. In addition, \kratos{}
framework also includes a thermochemistry module
\citep{2025arXiv250404941W}, which is adapted for
thermonuclear reactions in order to provide independent
benchmarks for us to compare and verify the evolution
results using \code{DeePODE}. In the calibration procedure presented in this work, we synchronize the hydrodynamic and nuclear reaction timesteps by setting both equal to the neural network timestep 
$\Delta t$. The simulation proceeds by performing one hydrodynamic step followed directly by one nuclear reaction step. \added{By employing this operator-splitting scheme, first-order temporal accuracy is maintained in the evolution of reacting flows.}

\subsection{Zero-dimensional calibration}
\label{sec:results-0d}

This section details zero-dimensional multi-step calibration for two networks, with extended multi-point single-step verification relegated to Appendix \ref{sec:appdx-tst-0d}. For \NetA, we conduct simulations using two different sets
of initial parameters, each run for 2,000 time steps. The
initial temperatures, densities, and mass fractions of
\{$^{4}$He,$^{12}$C,$^{16}$O\} for the two simulations are shown in Table \ref{table:0d-NetA}.

\begin{deluxetable}{ccc}
\label{table:0d-NetA}
\tabletypesize{\small} 
\tablewidth{0.45\textwidth} 
\tablecaption{Initial conditions of zero-dimensional calibrations for \NetA}
\tablehead{
  \colhead{Variables} & \colhead{\makecell{Calibration \Rmnum{1} \\ (Figure \ref{Fig.0d_sp13_test01})}} & \colhead{\makecell{Calibration \Rmnum{2} \\ (Figure \ref{Fig.0d_sp13_test02})}} 
}
\startdata
$T/\rm{(10^{8}~K)}$ & 4 & 4.5 \\
$\rho/\rm{(10^{6}~g~cm^{-3})}$  & 2& 1.1 \\
$X(^{4}\rm{He})$  & 0.6 & 0.4\\
$X(^{12}\rm{C})$  & 0.2 & 0.3\\
$X(^{16}\rm{O})$  & 0.2 & 0.3\\
\enddata
\end{deluxetable}

\begin{figure*}[htbp]
\centering  
\subfigure[]{
\label{Fig.0d_sp3_test01}
\includegraphics[width=0.45\textwidth]{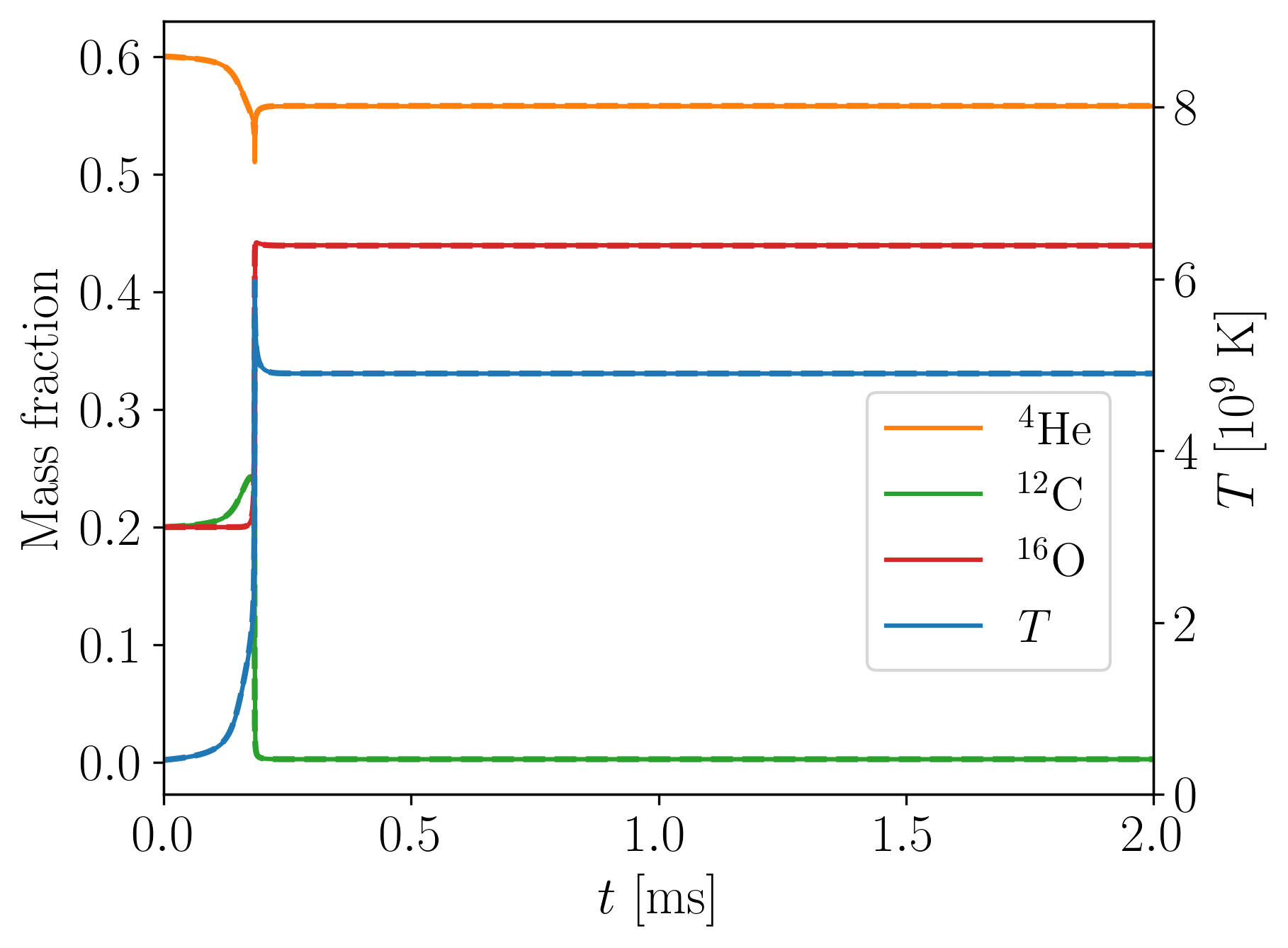}}
\subfigure[]{
\label{Fig.0d_sp3_test02}
\includegraphics[width=0.45\textwidth]{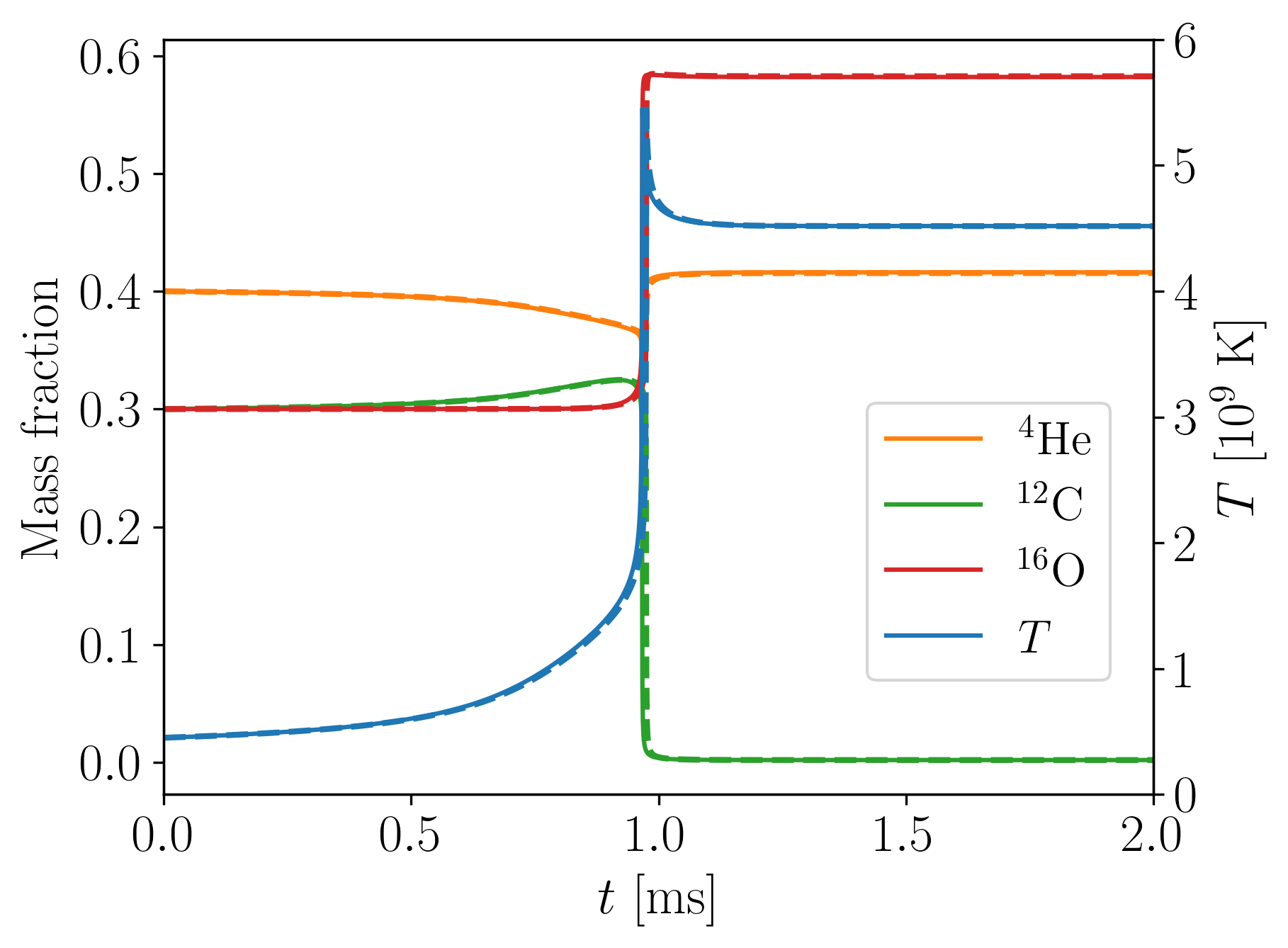}}
\caption{Two zero-dimensional calibrations for \NetA. We compare the temperature and mass fractions of various isotopes between neural network solutions (dashed lines) and direction integration solutions (solid lines) in simulations of 2 ms each. (a) Zero-dimensional calibration \Rmnum{1} for \NetA. (b) Zero-dimensional calibration \Rmnum{2} for \NetA.}
\label{Fig.0d_sp3}
\end{figure*}

The simulation results are presented in Figure \ref{Fig.0d_sp3}. In Figure \ref{Fig.0d_sp3_test01}, we observe that the results obtained from the neural network closely match those from the direction integration, with minimal differences.
In Figure \ref{Fig.0d_sp3_test02}, the results show only a slight inaccuracy in regions with high rigidity, while the rest of the results align very well with expectations. This minor discrepancy may arise from limitations described by the frequency principle \citep{luo_theory_2019}. However, this deviation has a negligible influence on the subsequent equilibrium state.

\begin{figure*}[htbp]
\centering  
\subfigure[]{
\label{Fig.0d_sp13_test01}
\includegraphics[width=0.45\textwidth]{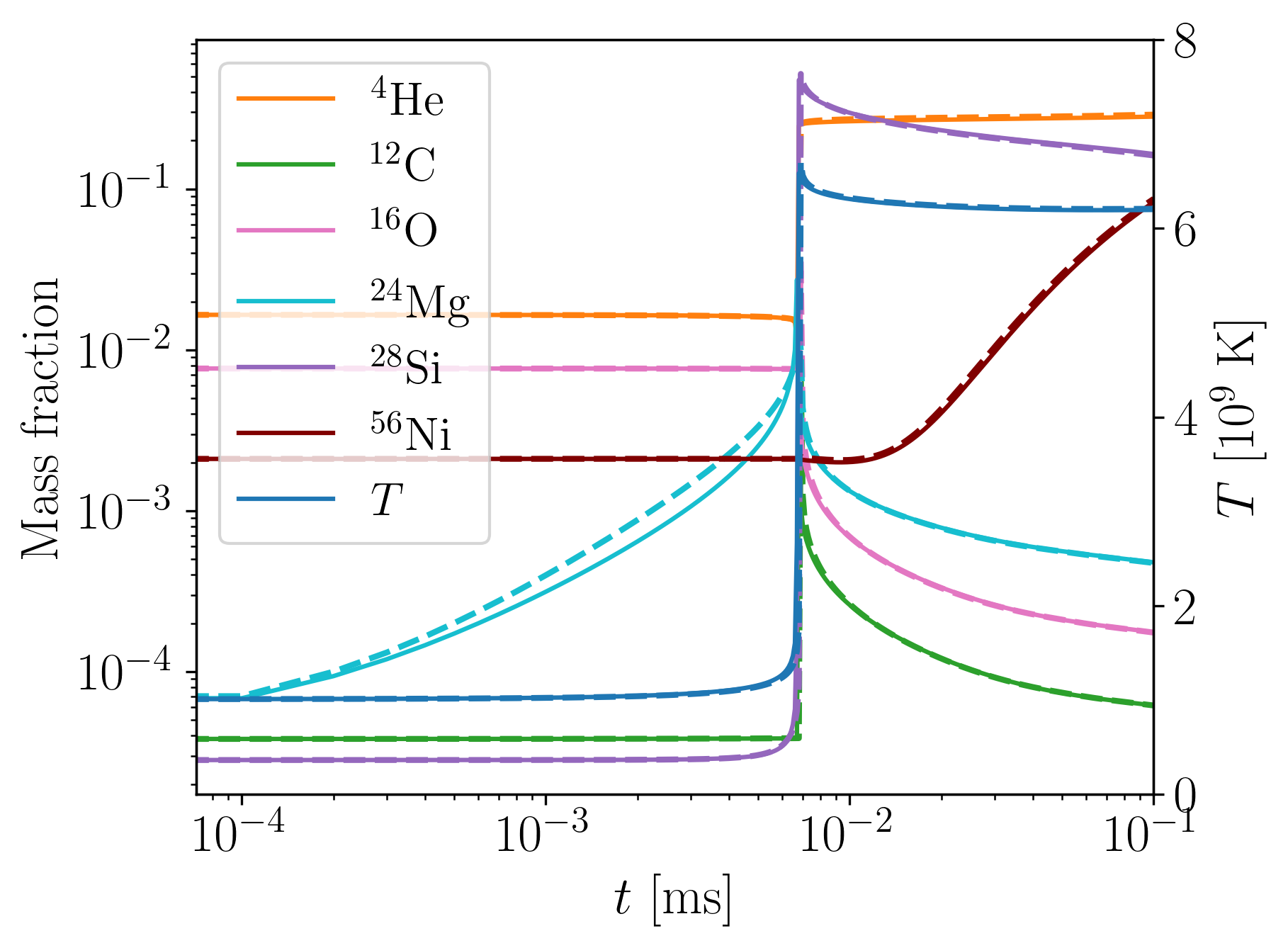}}
\subfigure[]{
\label{Fig.0d_sp13_test02}
\includegraphics[width=0.45\textwidth]{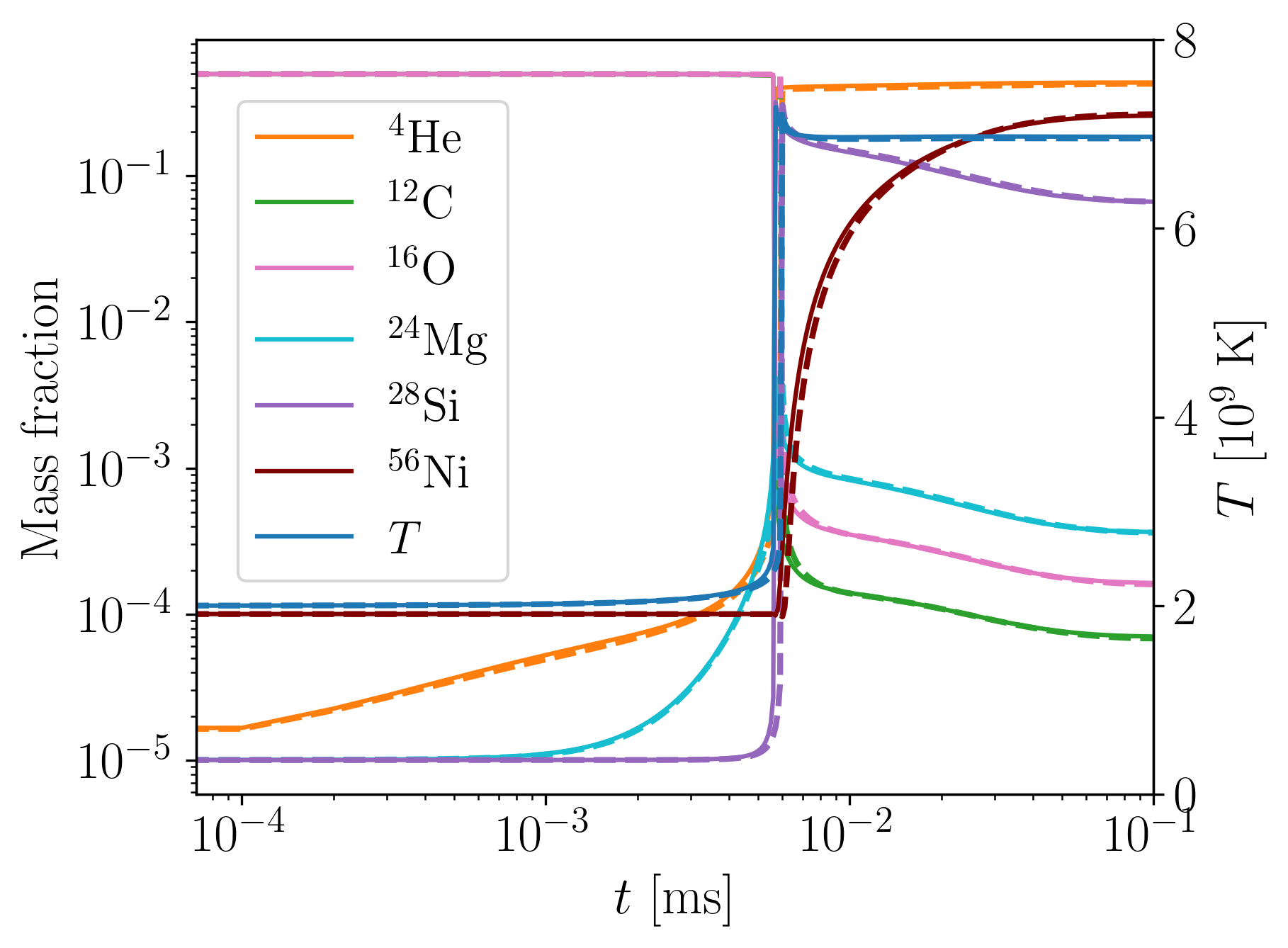}}
\caption{Two zero-dimensional calibration for \NetB. We compare the temperature and mass fractions of various isotopes between neural network solutions (dashed lines) and direction integration solutions (solid lines) in simulations of 0.1 ms each. (a) Zero-dimensional calibration \Rmnum{1} for \NetB. (b) Zero-dimensional calibration \Rmnum{2} for \NetB.}
\label{Fig.0d_sp13}
\end{figure*}

For \NetB, we perform simulations with two distinct sets of initial parameters, each run for 1,000 time steps. The initial parameters for the first case are randomly generated. For the second case, the parameters are derived from the conditions of internal deflagration in white dwarfs \citep{hristov_magnetohydrodynamical_2018}, with the following values: an initial temperature of $2\times10^{9}~\rm{K}$, an initial density of $1\times10^{8}~\rm{g}~\rm{cm}^{-3}$, and an initial isotope composition consisting of approximately 50\% $^{12}$C and 50\% $^{16}$O. The results, shown in Figure \ref{Fig.0d_sp13}, indicate that, apart from minor deviations in the mass fractions of certain isotopes during the steps of high rigidity, the overall agreement between the neural network and the direction integration is excellent.

\subsection{One-dimensional calibration}
\label{sec:results-1d}

\begin{figure*}[htbp]
\centering  
\subfigure[]{
\label{Fig.1d_sp3_cloud}
\includegraphics[width=0.45\textwidth]{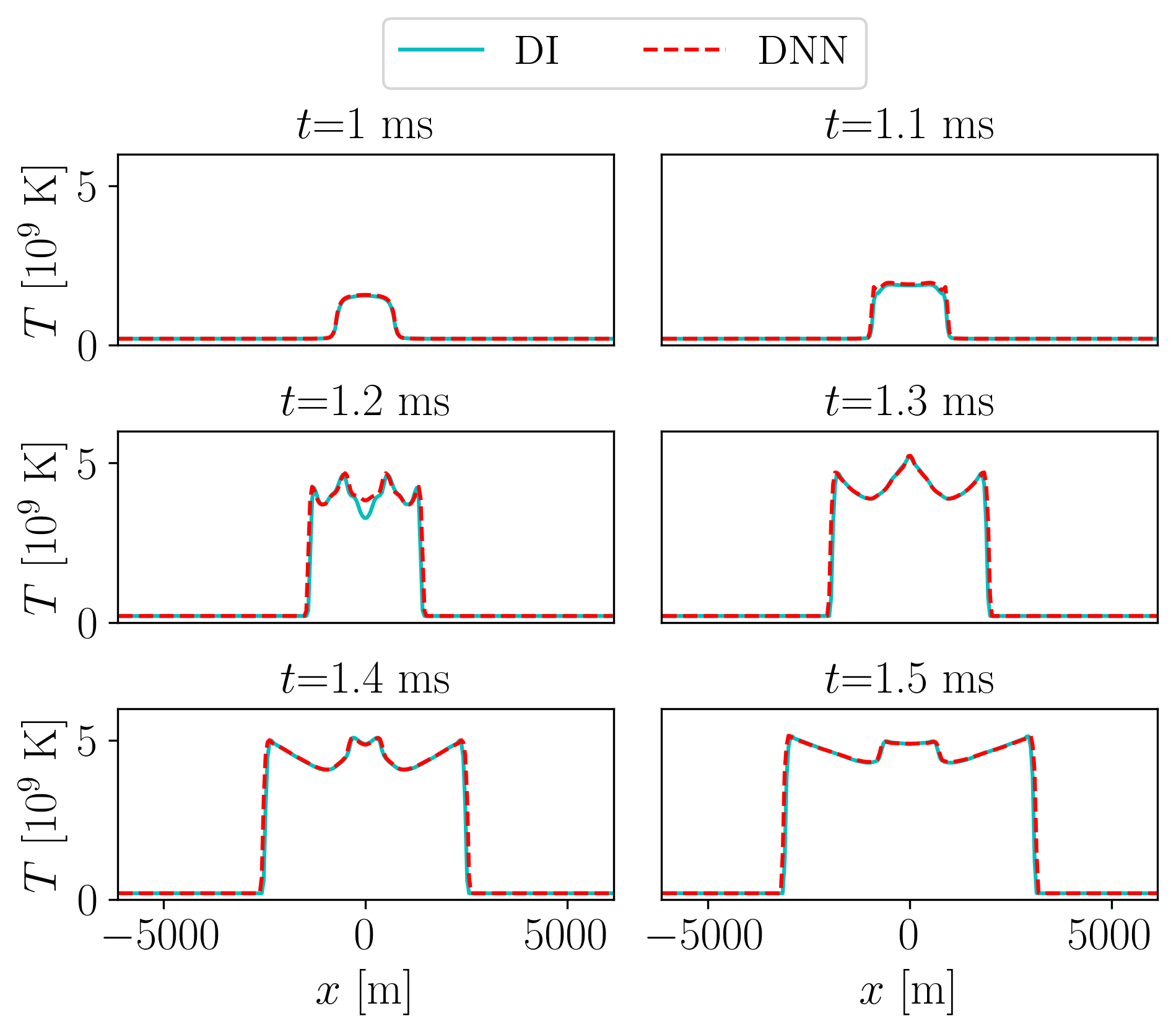}}
\subfigure[]{
\label{Fig.1d_sp3_compare}
\includegraphics[width=0.45\textwidth]{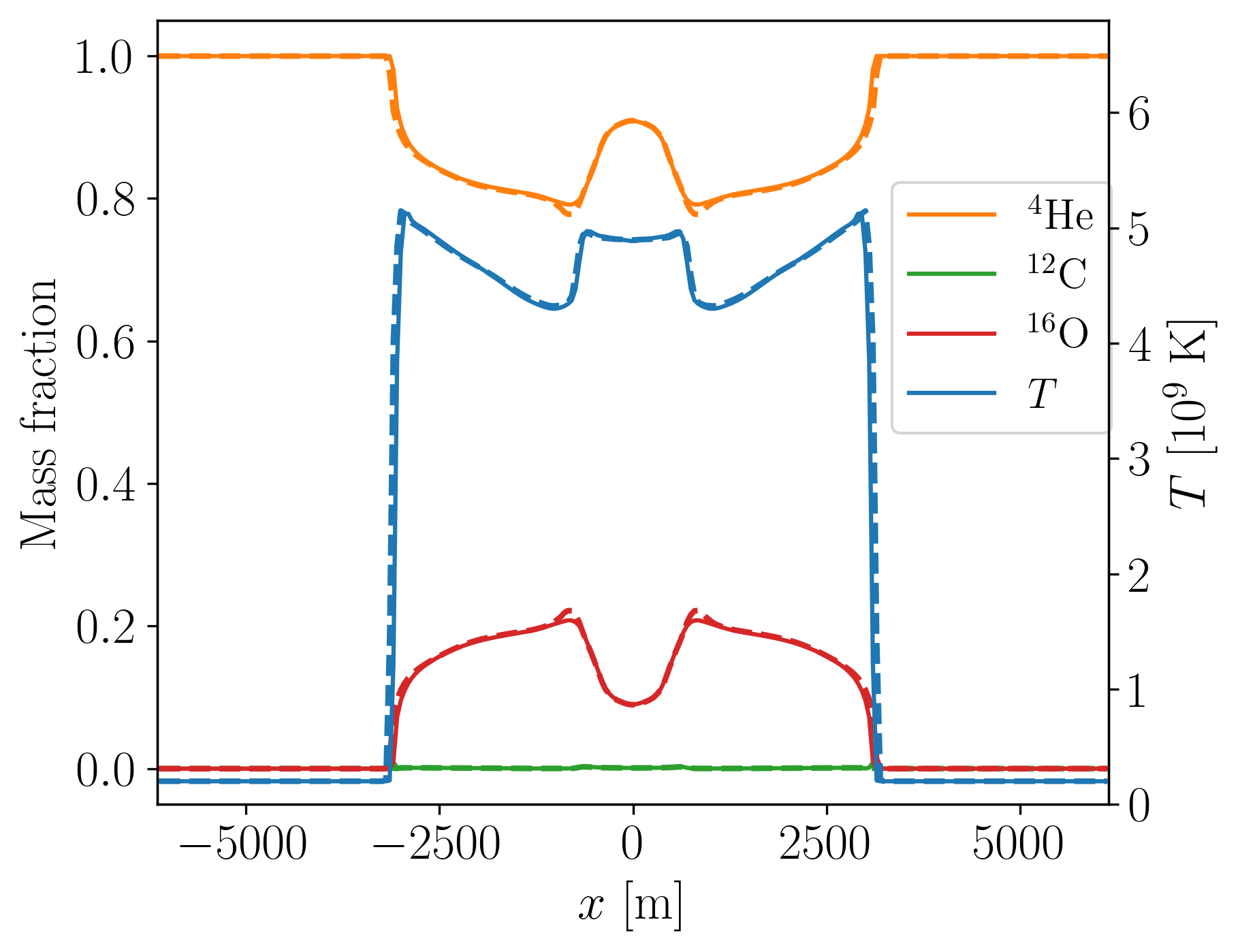}}
\caption{The one-dimensional simulation of \NetA. (a) The temperature change diagram of the one-dimensional case. (b) Comparison of temperature and mass fractions of isotopes obtained from simulations using neural networks (dashed lines) and direction integration (solid lines) at $t=1.5$ ms.}
\label{Fig.1d_sp3}
\end{figure*}

\begin{figure*}[htbp]
\centering  
\subfigure[]{
\label{Fig.1d_sp13_cloud}
\includegraphics[width=0.45\textwidth]{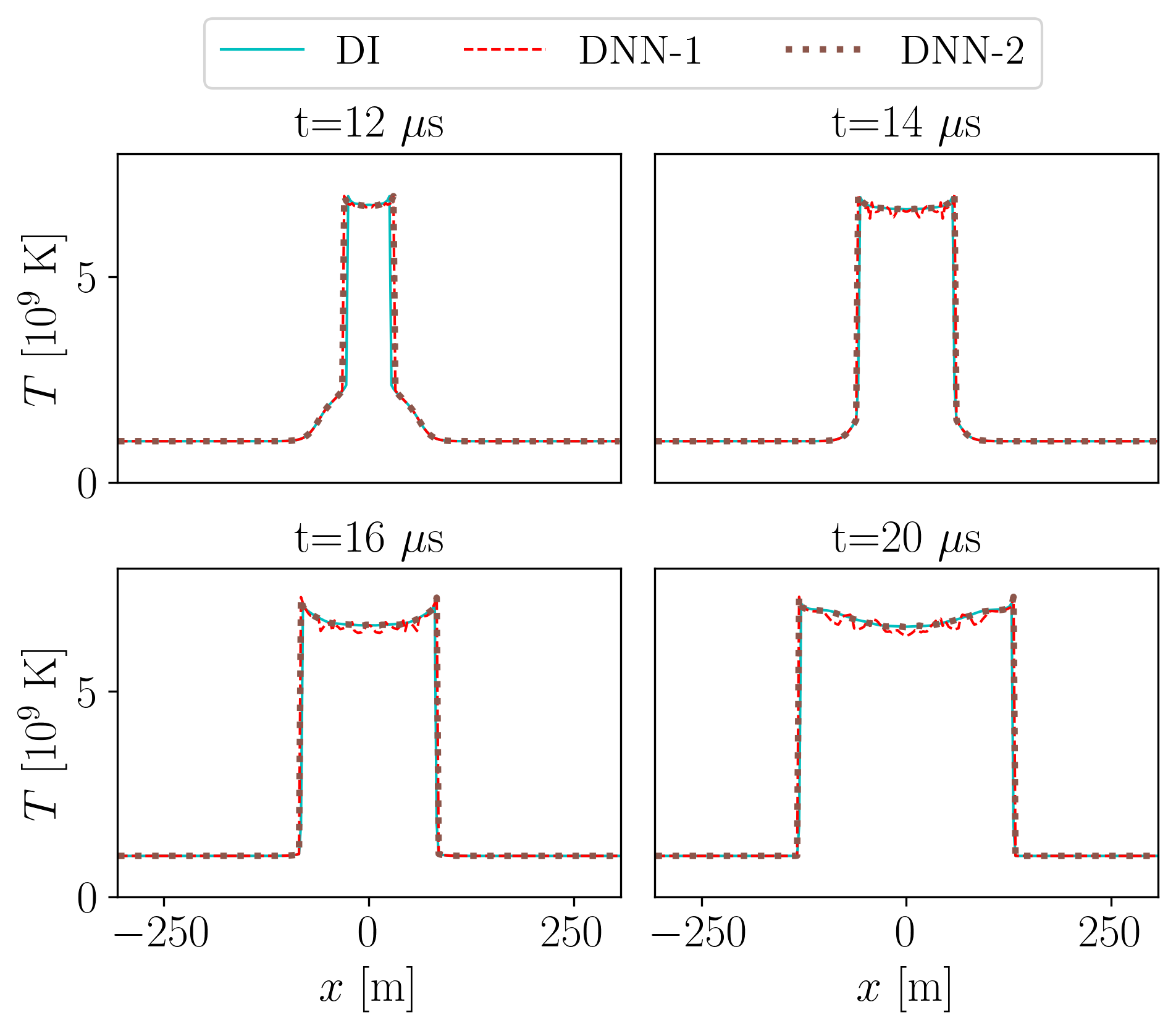}}
\subfigure[]{
\label{Fig.1d_sp13_compare}
\includegraphics[width=0.45\textwidth]{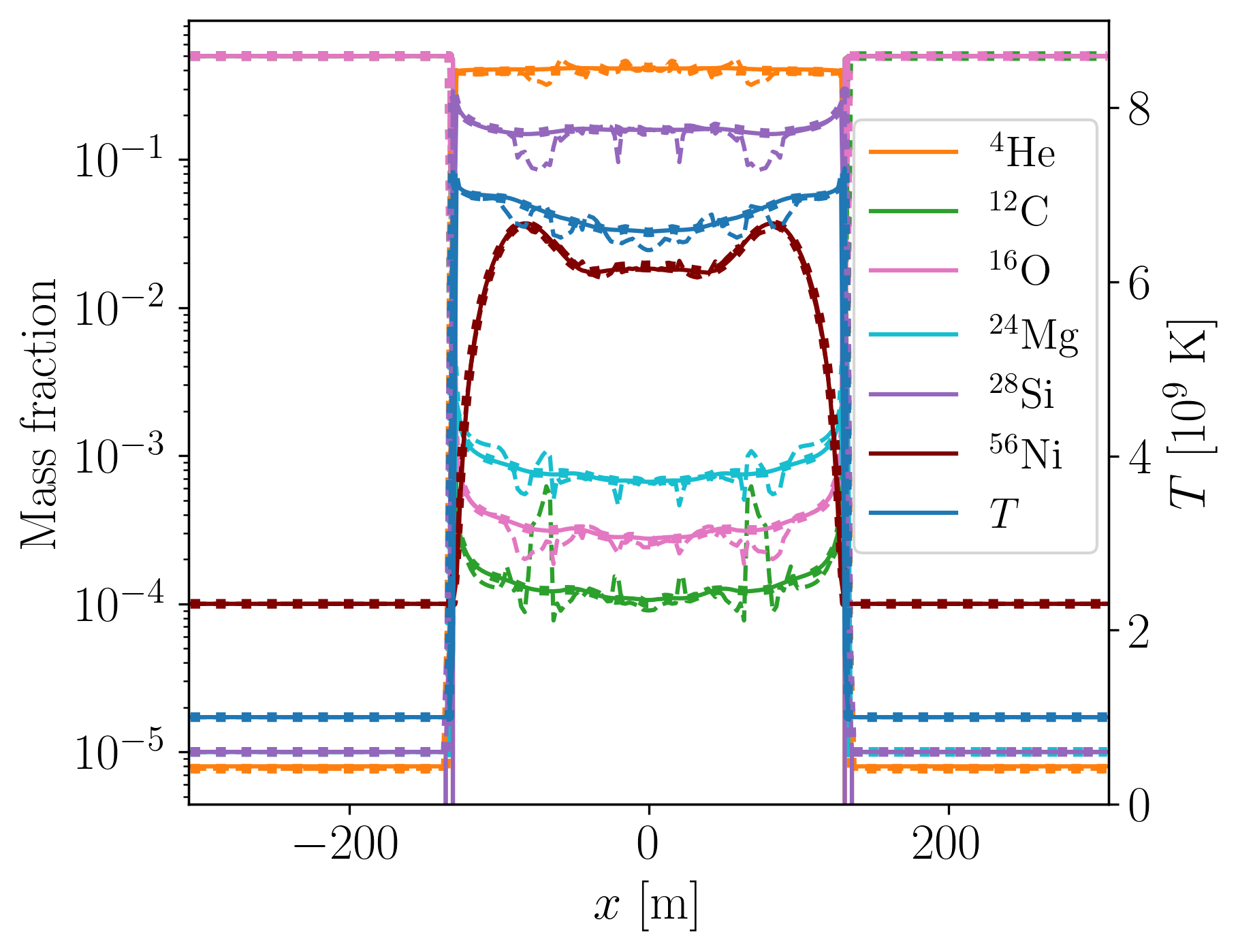}}
\caption{The one-dimensional simulation of \NetB. (a) The temperature change diagram of the one-dimensional case.  (b) Comparison of temperature and mass fractions of isotopes obtained from simulations using DNN-1 (dashed lines), DNN-2 (dotted lines) and direction integration (solid lines) at $t=20$ \textmu s.}
\label{Fig.1d_sp13}
\end{figure*}

For the one-dimensional calibration of \NetA, we adopt parameter settings corresponding to
surface combustion on white dwarfs during the double
detonation process described by
\citet{rivas_impact_2022}. The spatial domain of the
simulation spans $x/(10^{3}~\text{m})\in [-6.144, 6.144]$ with 256 grid cells and the boundary conditions are
outflow. The background conditions are as follows:
temperature $T_{0}=2\times 10^{8}~\rm{K}$, density
$\rho_{0}=1\times 10^{7}~\rm{g}~\rm{cm}^{-3}$. To
initiate the simulation, the temperature in the central
region is increased while maintaining constant pressure. The
new temperature is defined as:

\begin{equation}\label{1d_Tempterature}
    \begin{aligned}
        T=\frac{3T_{0}}{2}+\frac{T_{0}}{2}\tanh\left(4-\frac{|x|}{r^{\prime}}\right)\ ,
    \end{aligned}
\end{equation}
where $x$ represents the position of the grid cells, and $r^{\prime}=1.536\times10^{2}~\rm{m}$. The simulation runs for $1.5~\rm{ms}$. As shown in Figure \ref{Fig.1d_sp3_cloud}, the central region ignites after being preheated and subsequently propagates outward in both directions. The results, presented in Figure \ref{Fig.1d_sp13_compare}, demonstrate excellent agreement between the neural network and the direction integration, with only minor deviations in the mass fraction observed in a few grid cells.

For \NetB, we utilize parameter configurations
representative of internal deflagration in white dwarfs,
following the work of
\citet{hristov_magnetohydrodynamical_2018}. The spatial
domain of the simulation is
$x/(10^{2}~\rm{m})\in[-3.072, 3.072]$ with 256 grid
cells, the boundary conditions are also outflow. The
background conditions are as follows: temperature $T_{0}=10^{9}~\rm{K}$, density
$\rho_{0}=10^{8}~\rm{g}~\rm{cm}^{-3}$
and the initial isotope composition consists of
approximately 50\% $^{12}$C and 50\% $^{16}$O. As with \NetA, a controlled thermal perturbation is imposed within the center domain while
maintaining constant pressure. The temperature distribution
is the same as Equation \ref{1d_Tempterature} but with
$r^{\prime}=15.36~ \rm m $. And this simulation runs for $20~{\rm \mu s}$.

Figure \ref{Fig.1d_sp13_cloud} shows the preheated core igniting with bidirectional flame propagation, the same as Figure \ref{Fig.1d_sp3_cloud}. The results, illustrated in Figure \ref{Fig.1d_sp13_compare}, show that when the simulation uses the neural network entirely (DNN-1), the reacted region exhibits small deviation compared to the direction integration, and oscillations are observed. This indicates limitations in the neural network's accuracy near the reaction equilibrium region. Consequently, we set the upper temperature limit of the neural network to $6.5\times10^{9}$ K (DNN-2), where the computational results exhibit close alignment with direct numerical integration. Under these operational conditions, the neural network's computational utilization rate is maintained above 88\%, demonstrating its efficiency. This strategy can be useful in practice, since direct integration is deemed sufficient for treating the equilibrium region, while the neural network tends to outperform in stiffer ones. In addition, incorporating the nuclear statistical equilibrium detection method mentioned in \cite{kushnir_accurate_2020} may also provide an effective solution for handling the equilibrium region.

\subsection{Two-dimensional calibration}
\label{sec:results-2d}

\begin{figure*}[htbp]
\centering  
\subfigure[]{
\label{Fig.2d_sp3_cloud}
\includegraphics[width=0.46\textwidth]{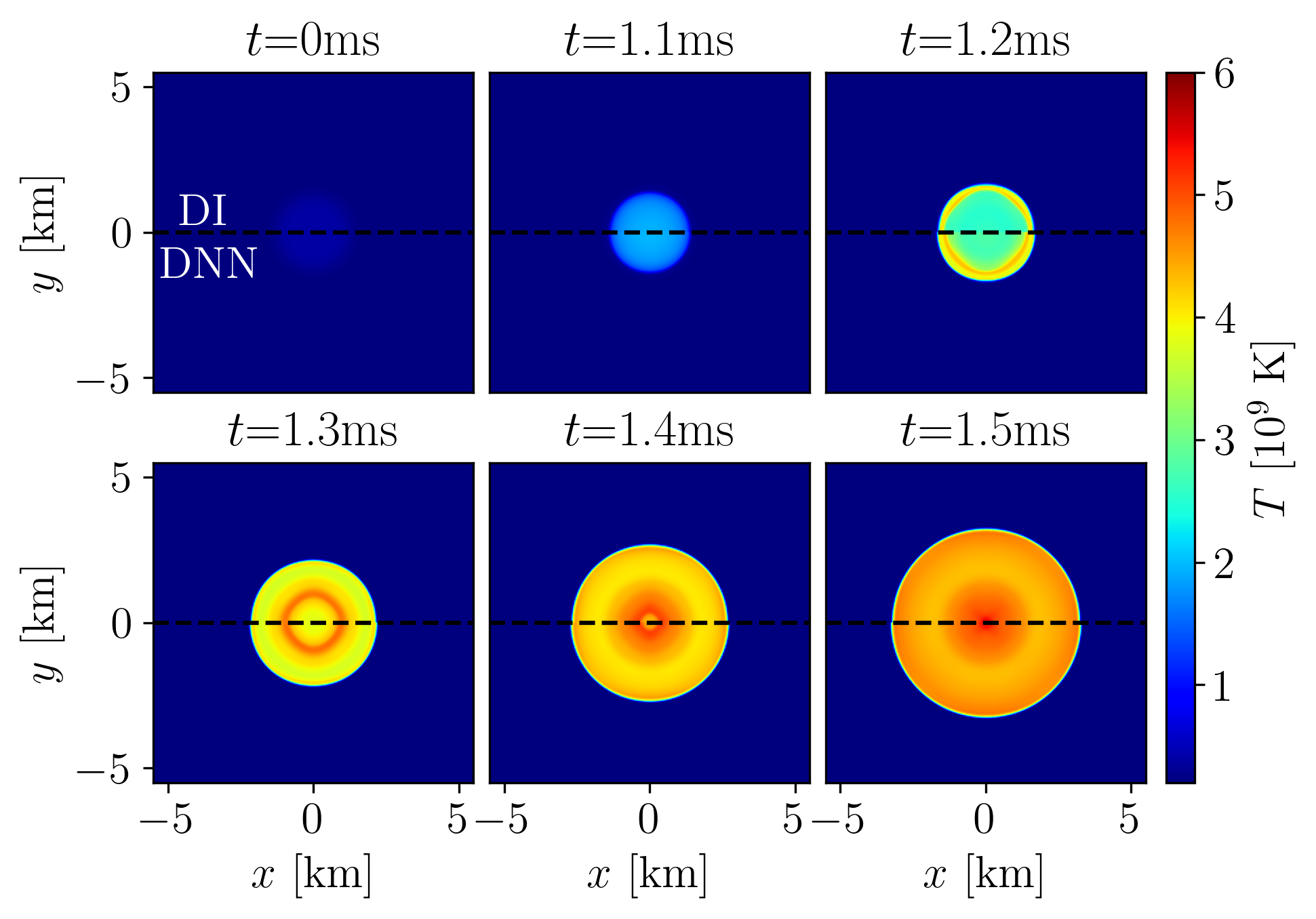}}
\subfigure[]{
\label{Fig.2d_sp3_compare}
\includegraphics[width=0.45\textwidth]{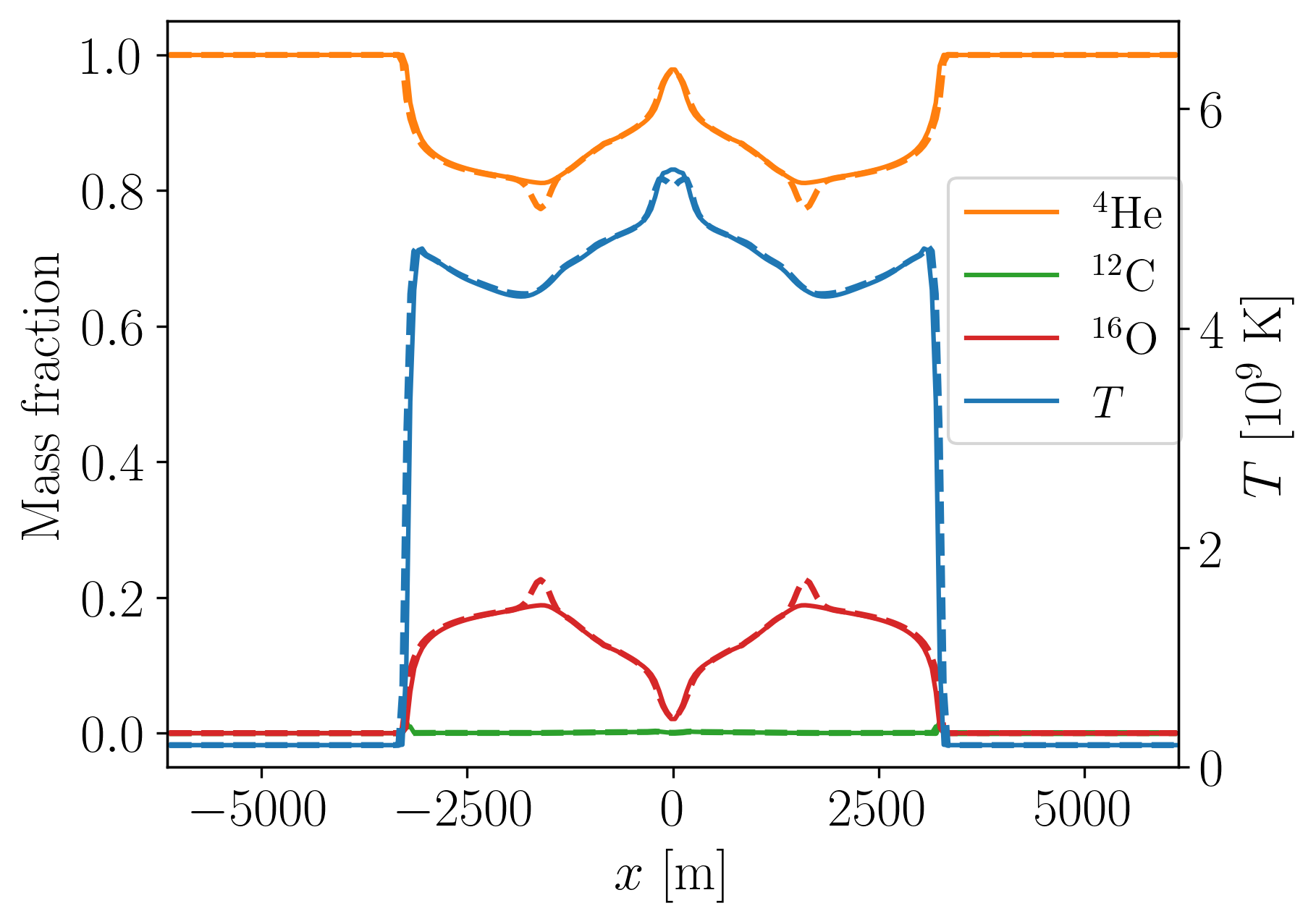}}
\caption{Two-dimensional simulations using \NetA. (a) The
  upper and lower halves of the subplots show the
  temperature evolution from the direction integration and
  the neural network, respectively. (b) Comparisons of the
  temperature and mass fractions of isotopes obtained from
  simulations using the neural network (dashed lines) and
  the direct integration (solid lines) at
  $t=1.5\ \rm{ms}$ along the $y=0$ axis.}
\label{Fig.2d_sp3}
\end{figure*}

The first two-dimensional calibration for \NetA\ extends the
one-dimensional setup to a two-dimensional spatial domain:
$x/(10^{3}~\text{m})\in[-6.144, 6.144]$, $y/(10^{3} ~\text{m})\in[-6.144, 6.144]$ with a grid
resolution of 256$\times$256 cells, the boundary conditions
in the x and y directions are both outflow. The background
conditions are defined as follows: temperature
$T_{0}=2\times 10^{8}\ \rm{K}$, density
$\rho_{0}=1\times 10^{7}~\rm{g}~ \rm{cm}^{-3}$, and
isotopic composition is almost entirely composed of
$^{4}$He. The perturbation configuration retains structural consistency with the one-dimensional calibration \eqref{1d_Tempterature}, with the spatial coordinate systematically substituted from $x$ to $R=\sqrt{x^{2}+y^{2}}$, and $r^{\prime}=3.072\times10^{2}$ m. This simulation
also runs for $1.5~\rm{ms}$.

After the simulation, the results are analyzed by extracting data points along the line $y=0$ for comparison. As shown in the  Figure \ref{Fig.2d_sp3}, the temperature and density profiles, as well as the mass fractions, align closely with those obtained using the direction integration. Deviations are observed in the mass fractions for only a small number of grid cells, similar to the one-dimensional calibration results. These discrepancies are minimal and do not significantly affect the overall accuracy of the simulation.

\begin{figure*}[htbp]
\centering  
\includegraphics[width=1.0\textwidth]{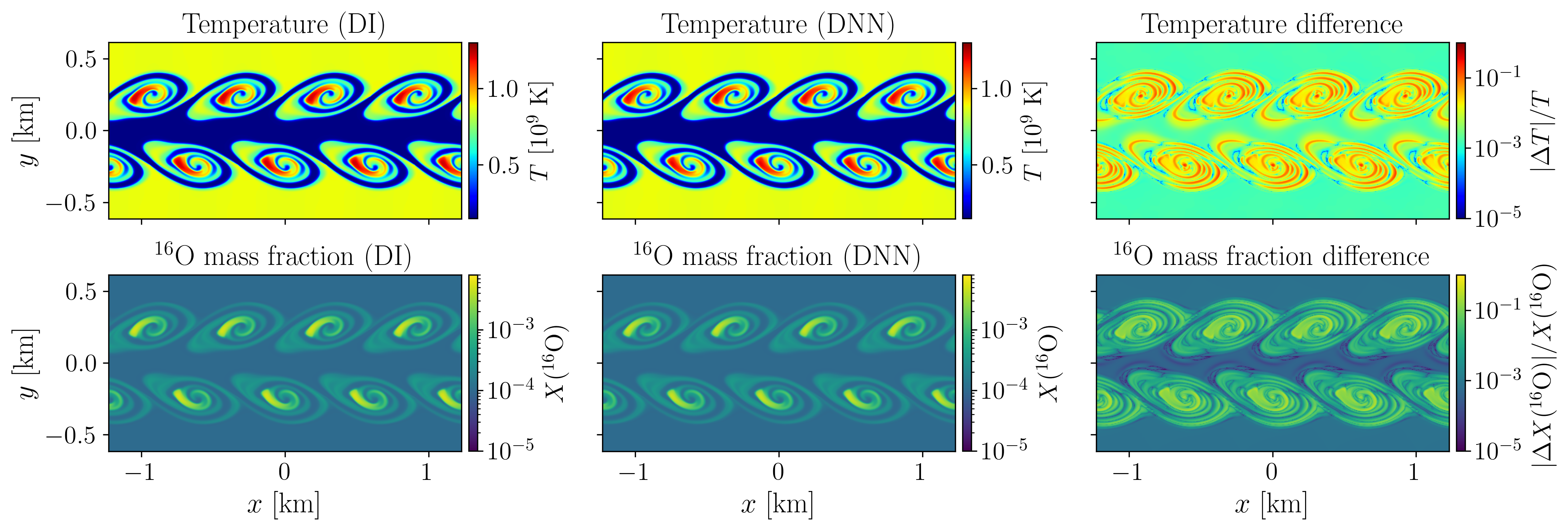}
\caption{The Kelvin-Helmholtz instability tests results, comparing the temperature (upper row) and mass fraction of O16 (lower row) at $t = 6.5\ \rm{ms}$ from the direction integration (left column) and the DNN (middle column) results. The right column shows the relative difference.}
\label{Fig.KHI_s3}
\end{figure*}

The second
two-dimensional calibration for \NetA\ is the
Kelvin-Helmholtz instabilities case, which is motivated by and adapted from
\citet{casanova_kelvinhelmholtz_2011}. The spatial domain is
defined as: $x/(10^{3}~\text{m})\in[-1.2288, 1.2288]$, $y/(10^{2}~\text{m})\in[-6.144, 6.144]$ with a
grid resolution of 512$\times$256 cells. The
  $x$-boundary conditions are periodic, while the
  $y$-boundary conditions are reflecting. The initial
conditions are set as:
\begin{align}
    &\rho = \rho_{0}+\delta\rho \tanh\left(\frac{y_{0}-|y|}{L}\right)\ , \\
    &v_{x} = v_{x0}\tanh\left(\frac{y_{0}-|y|}{L}\right)\ , \\
    &v_{y} = v_{y0}\cos\left(\frac{2\pi}{\lambda}\right)\exp\left[-\frac{(|y|-y_{0})^{2}}{2y_{0}^{2}}\right]\ , \\
    &X(^{4}\text{He}) = X_{0} + \frac{1-3X_{0}}{2}\left[1+\tanh\left(\frac{y_{0}-|y|}{L}\right)\right]\ , \\
    &X(^{12}\text{C}) = X_{0} + \frac{1-3X_{0}}{2}\left[1-\tanh\left(\frac{y_{0}-|y|}{L}\right)\right]\ , \\
    &X(^{16}\text{O}) = X_{0}\ ,
\end{align}
where $\rho_{0}$, $X_{0}$, $v_{x0}$ and $v_{y0}$ are the
reference values for density, mass fraction and $x$, $y$
velocity components, $\delta\rho$ is the difference in
density across the shearing layer with thickness $L$,
$\lambda$ is the velocity perturbation wavelength. Here,
$\rho_{0} = 1.5\times10^{7}~\rm{g}~\rm{cm}^{-3}$,
$\delta\rho = 4.5\times10^{6}~\rm{g}~\rm{cm}^{-3}$,
$y_{0} = 2.048\times10^{2}~\rm{m}$, $L = 30.72~\rm{m}$,
$v_{x0} = -2\times10^{7}~\rm{cm}\  \rm{s}^{-1}$,
$v_{x0} = 2\times10^{6}~\rm{cm}\ \rm{s}^{-1}$,
$\lambda = 6.144\times10^{2}~\rm{m}$, $X_0 = 10^{-4}$.

The results are shown in Figure \ref{Fig.KHI_s3}, which compares the temperature and the mass fraction of $^{16}$O at $t=6.5~\rm{ms}$. As evidenced by the figure, both temperature profiles and $^{16}$O mass fractions exhibit minimal variations, demonstrating maximum relative errors within approximately 20\%. The observed discrepancies are predominantly localized near the flow boundary regions, potentially attributable to minor structural displacements during the transport processes.

\begin{figure*}[htbp]
\centering  
\includegraphics[width=0.9\textwidth]{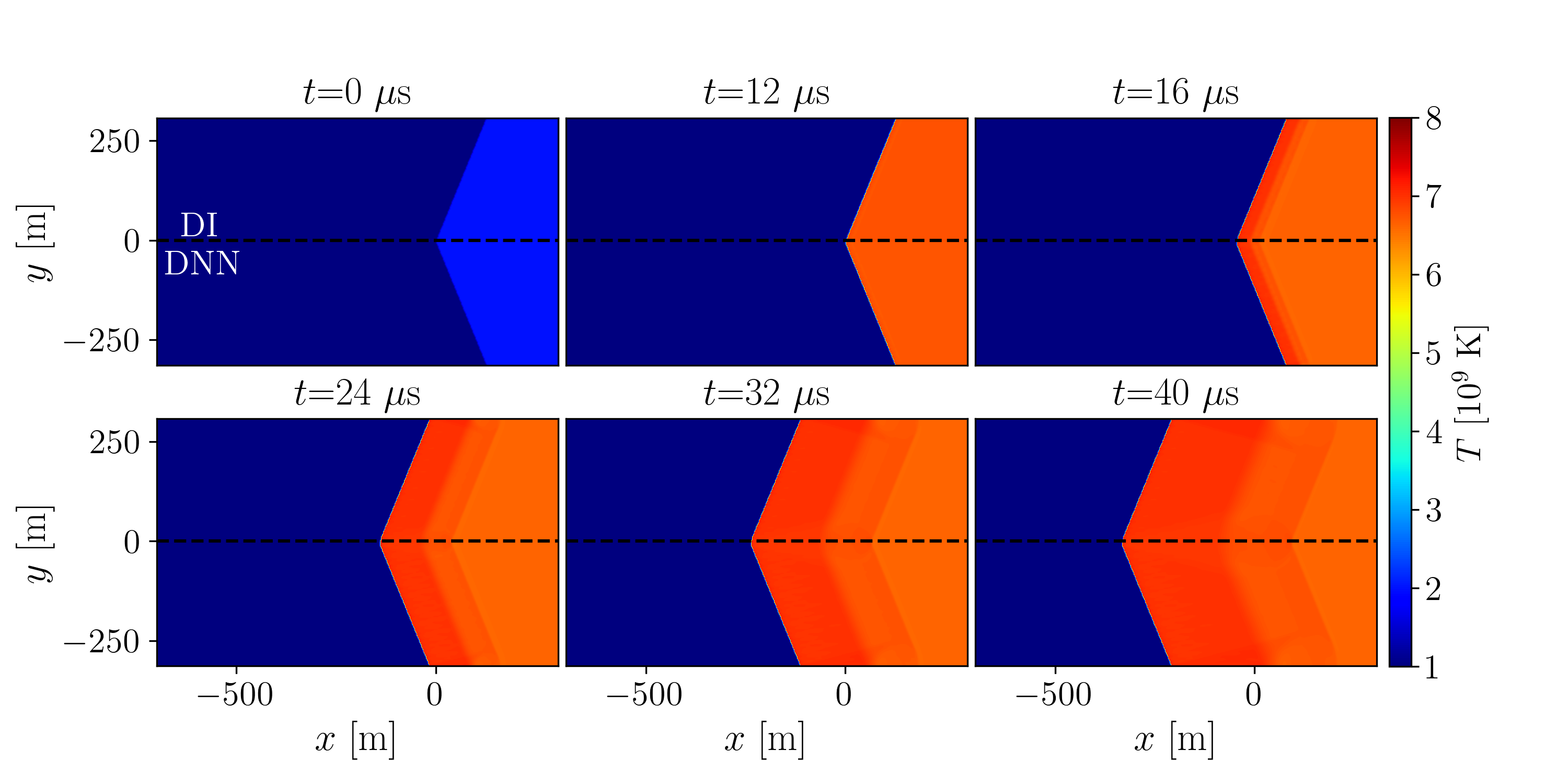}
\caption{The temperature change diagram of the two-dimensional case for \NetB. The upper part of the subplot in each step shows the direction integration results, and the lower part shows the neural network results.}
\label{Fig.2d_sp13_cloud}
\end{figure*}

\begin{figure}[htbp]
\centering  
\includegraphics[width=0.45\textwidth]{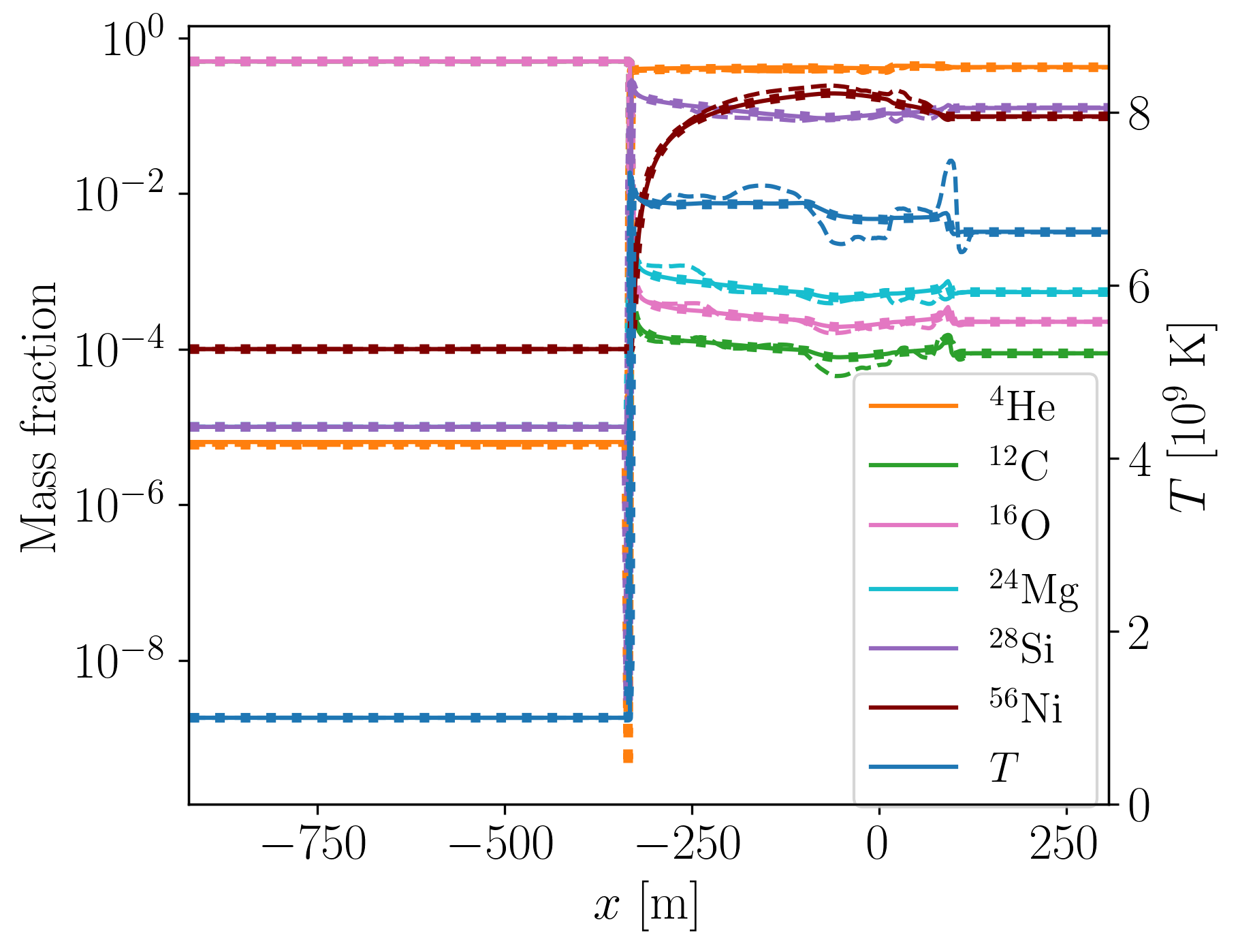}
\caption{Comparison of temperature and mass fractions of isotopes obtained from two-dimensional simulations of \NetB{} using DNN-1 (dashed lines), DNN-2 (dotted lines) and direction integration (solid lines) at $t=40~{\rm \mu s}$
 along the axis where $y=0$.}
\label{Fig.2d_sp13_compare}
\end{figure}

For the two-dimensional calibration of \NetB, we employ a V-shaped flame configuration (Figures \ref{Fig.2d_sp13_cloud}, \ref{Fig.2d_sp13_compare}), which also utilizes parameter configurations derived from \citet{hristov_magnetohydrodynamical_2018}. The spatial domain is defined as: $x/(10^{2}~\rm{m})\in[-9.216, 3.072]$, $y/(10^{2}~\rm{m})\in[-3.072, 3.072]$ with a grid resolution of 512$\times$256 cells, the x-boundary conditions are outflow and reflect, the y-boundary conditions are periodic. The initial conditions are as follows:

\begin{itemize}
\item[$\bullet$] Upstream: $T=2\times10^{9}~\rm{K}$, $\rho=5\times10^{7}~\rm{g}~\rm{cm}^{-3}$
\item[$\bullet$] Downstream: $T=1\times10^{9}~\rm{K}$, $\rho=1\times10^{8}~\rm{g}~ \rm{cm}^{-3}$
\item[$\bullet$]Isotopic composition (both regions): approximately $50\%$ $^{12}$C and $50\%$ $^{16}$O.
\end{itemize}

This simulation runs for $40~{\rm \mu s}$. During the simulation,
the flame on the right ignites after preheating and
propagates towards the left, as shown in Figure
\ref{Fig.2d_sp13_cloud}. Same as the 1D calibration, two simulations are performed: full adoption of the neural network (DNN-1) and temperature-limited adoption below $6.5\times10^{9}~\rm{K}$ (DNN-2), with identical grid parameters. After the
simulations, data points along the line $y=0$ are extracted
for comparison. As demonstrated in Figure \ref{Fig.2d_sp13_compare}, when the neural network operates continuously (DNN-1) during the simulation, the reaction zone exhibits minor deviations accompanied by pronounced oscillations. However, implementing a temperature activation threshold (DNN-2) for the network yields results that demonstrate strong agreement with direct integration values. This adjustment significantly improves stability while preserving accuracy, and the network operates at 76\% utilization under thermal gating.

\subsection{Computational time cost comparison}
\label{sec:results-time-cost}

\begin{figure}[htbp]
\label{Fig.time_compare}
\includegraphics[width=0.45\textwidth]{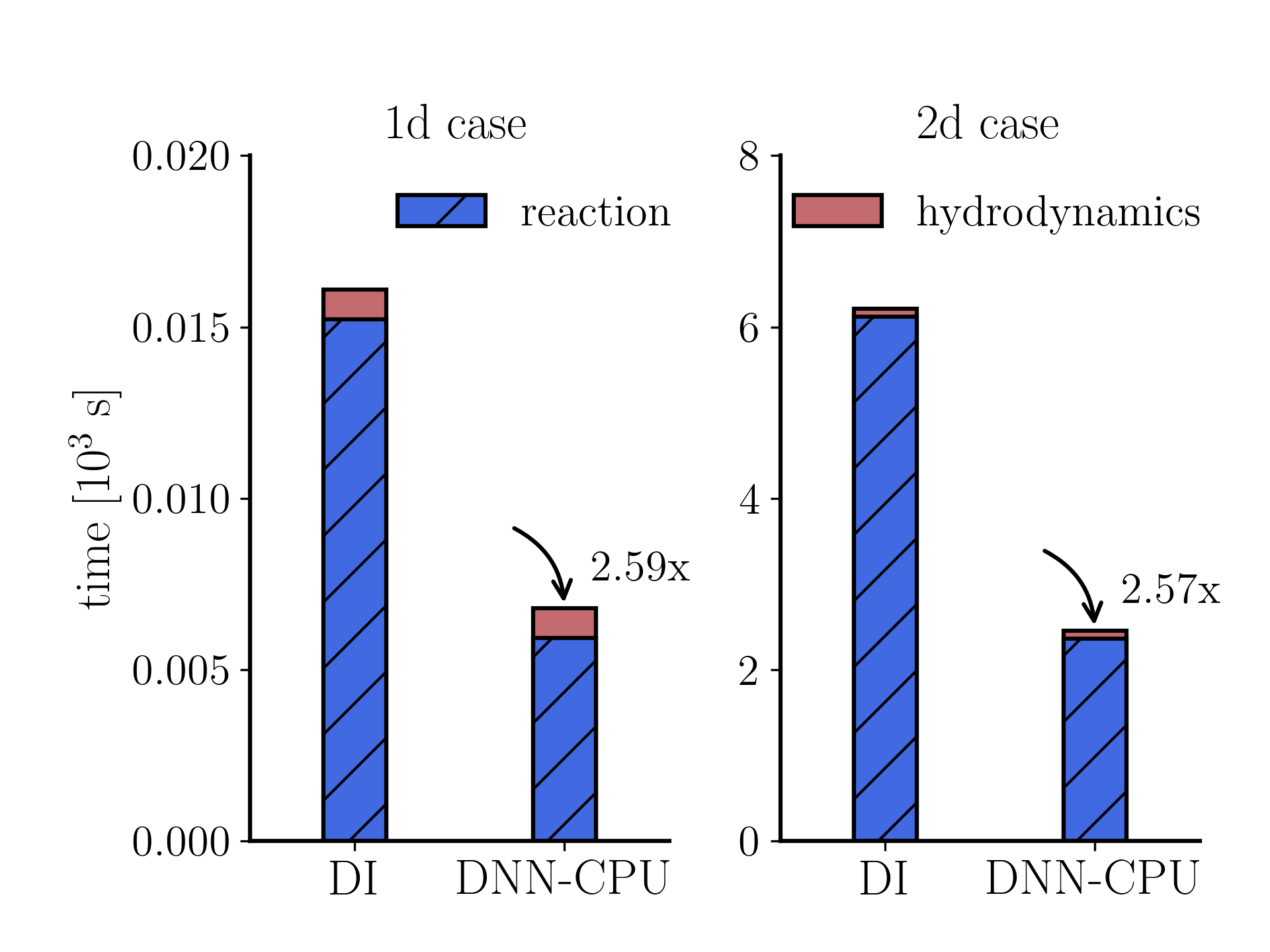}
\caption{The comparison of computation time between solving nuclear reactions using neural networks and direction integration. The left panel shows the one-dimensional case, while the right panel shows the two-dimensional case.}
\end{figure}

While GPU architectures are generally considered more suitable for
neural network acceleration due to their massive parallel
computing capabilities, our current implementation focuses
on CPU-based validation as a crucial first step
benchmark. Experimental results demonstrate that our neural
network integration achieves $\sim 2.59\times$ and $\sim 2.57\times$
(\ref{Fig.time_compare}) speedup for 13-species simulations
in 1D (Figure \ref{Fig.1d_sp13}) and 2D (Figures
\ref{Fig.2d_sp13_cloud}, \ref{Fig.2d_sp13_compare})
configurations respectively, when executed on a six-core
Intel i5-12400F CPU processor.

Notably, these CPU-based acceleration metrics should be interpreted as conservative estimates of the method's potential. The intrinsic parallelism of neural network computations remains fundamentally constrained by CPU architectures' limited thread scalability and memory bandwidth. To fully exploit the algorithm's acceleration capacity, our ongoing work involves integrating the neural network component directly into the GPU software architecture, rather than leveraging PyTorch for CPU-side invocation. This migration is expected to \rmnum{1}) establish more realistic acceleration baselines through explicit comparison of CPU/GPU computational graphs, and \rmnum{2}) leverage GPU tensor cores' mixed-precision computing to further optimize the inference latency-critical components.

\section{Summary and Future Works}
\label{sec:summary}

This work demonstrates the successful development and
validation of deep neural network (DNN) surrogate models for
accelerating astrophysical nuclear reaction simulations. By
leveraging the \code{DeePODE} framework, we established a surrogate
models for both 3-species and 13-species reaction networks
(\NetA{} and \NetB), achieving comparable accuracy to direct
integration while attaining a $\sim 2.6\times$ speedup on
CPU architectures. The models generalize robustly, across
zero-dimensional calculations (temporal evolution at one
spatial point) through two-dimensional simulations, via the
temperature-thresholded deployment strategies, maintaining
neural network utilization rates above 75\% in
multi-dimensional scenarios. With the evolutionary Monte
Carlo sampling (EMCS) employed for broad-phase-space data
generation, and Box-Cox transformations that handle
multi-scale mass fractions, the \code{DeePODE} module
perform fast and accurate predictions within the hybrid
integration paradigm, dynamically balancing the capability
of DNN predictions in stiff regions with traditional solvers
adopted elsewhere. Validation across flame propagation,
Kelvin-Helmholtz instabilities, and white dwarf deflagration
scenarios confirm the framework's ability to resolve
complex reacting flows, while accelerating the overall
computation by circumventing stiffness limitations inherent
to conventional ODE solvers.

While this paper establishes a foundation for data-driven
modeling of nuclear reacting flows, several critical
advancements remain to fully realize the potential of DNN
surrogates in astrophysical simulations. One possibility is
implementing on-the-fly training, where the neural network
is dynamically trained and updated during the initial stages
of a simulation. This paradigm would reduce the need for
precomputed datasets by integrating real-time data
generation and model refinement into the simulation
workflow. 
Expanding the scope to larger
reaction networks (e.g. $\sim 100$ species) and
multi-physics couplings
could also be promising in implementing 
these data-driven methods in extreme regimes with higher
complications. Coupling DNN surrogates with 
regression methods could yield interpretable, reduced-order models
for nucleosynthesis pathways, bridging machine learning with 
theoretical nuclear astrophysics.
These advancements could
collectively enable high-fidelity, end-to-end simulations of
multi-scale astrophysical phenomena, while overcoming the
computational barriers imposed by conventional stiff
solvers.
\\
\\
Code availability: As related manuscripts are still under review, the code for training and deploying the neural network is not yet publicly available. In the future, we plan to release both the code for training and utilizing the neural network, along with the pretrained models.
\\
\\
Software: The reaction networks are generated using the pynucastro \citep{smith_pynucastro_2023, pynucastro, the_pynucastro_development_2024_13899727}. The sampling process is implemented in the Castro hydrodynamics code \citep{zingale_improved_2019,Almgren2020,2010ApJ...715.1221A,the_castro_development_team_2025_14589933}, which is built upon the Starkiller Microphysics libraries \citep{amrex_astro_microphysics_development_tea_2025_14584895}, and the AMReX framework \citep{AMReX_JOSS,the_amrex_development_team_2025_14589566}.  The yt interface \citep{turk_yt_2011} are used to extract data from Castro. \kratos{} \citep{wang2025} is employed to perform the calibration of nuclear reacting flow simulations. Integration of the neural network into the \kratos{} environment is achieved through the PyTorch framework. \citep{NEURIPS2019_9015}.


\begin{acknowledgments}

\noindent This work is supported by the National Natural Science Foundation of China under Grants No. 92270001(Z. X.), 12371511 (Z. X.), 12422119 (Z. X.), 92470127 (T. Z.), and 12305246 (Y. Z.) and the Fundamental Research Funds for the Central Universities.

\end{acknowledgments}







\appendix

\section{Single-step tests with zero spatial dimensions}
\label{sec:appdx-tst-0d}

\begin{figure*}[htbp]
\centering  
\includegraphics[width=1.0\textwidth]{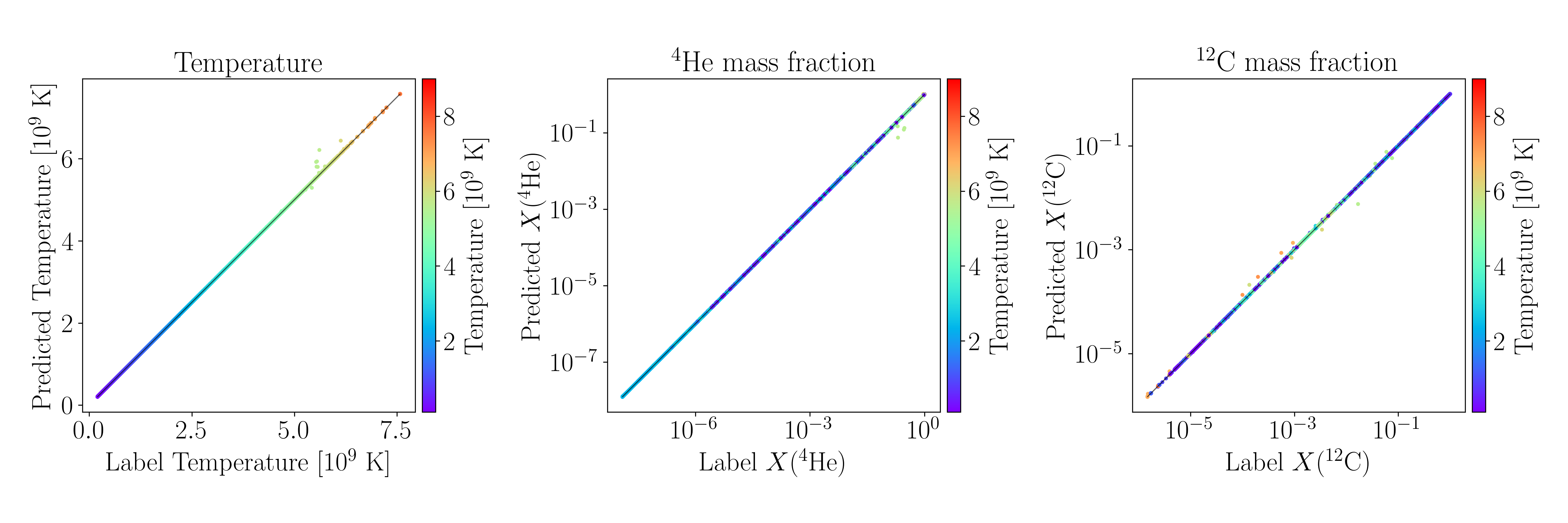}
\caption{One step tests for \NetA. The horizontal and
  vertical axes represent the results of one step predicted
  by the direction integration and the neural network,
  respectively.}
\label{Fig.0d_sp3_compare_T}
\end{figure*}

\begin{figure*}[htbp]
\centering  
\includegraphics[width=1.0\textwidth]{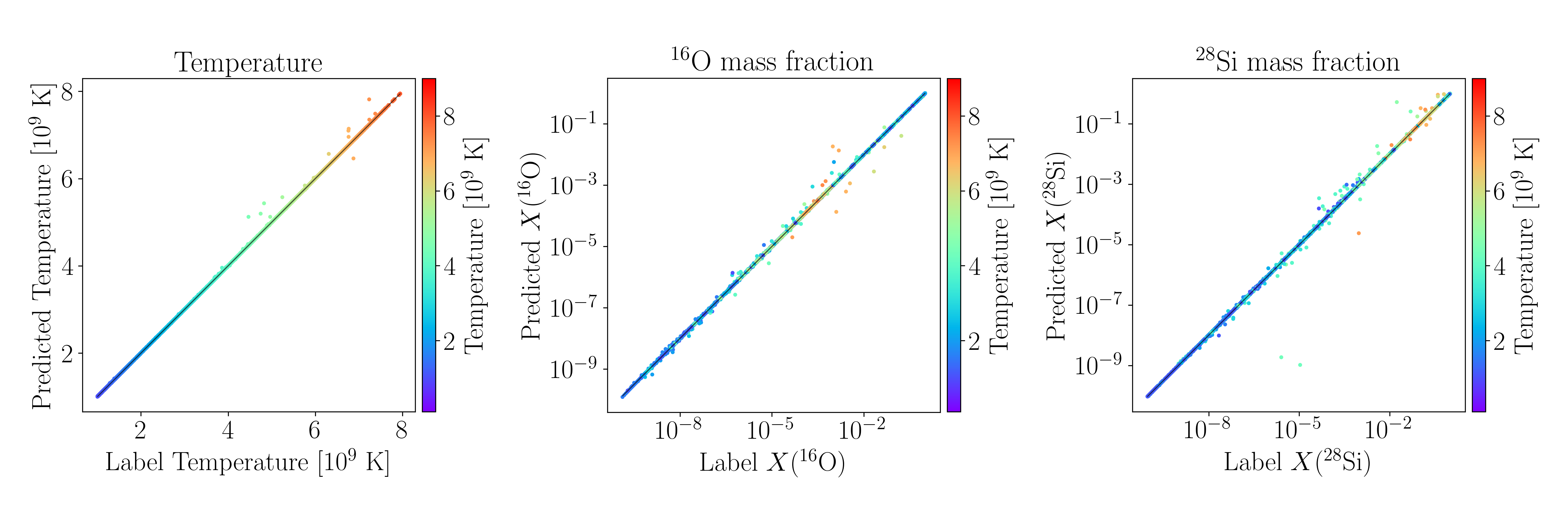}
\caption{One step tests for \NetB. The horizontal and vertical axes represent the results of one step predicted by the direction integration and the neural network, respectively.}
\label{Fig.0d_sp13_compare_T}
\end{figure*}

We conduct a large number of zero-dimensional single-step
tests and compare the results with those of direction
integration. The method of data collection is similar to
sampling, but these test data are not included in the
training set. Approximately $5\times 10^5$
sets of data are used in the testing of \NetA, and
approximately 1 million sets of data are used in the testing
of \NetB. From Figure \ref{Fig.0d_sp3_compare_T} and Figure
\ref{Fig.0d_sp13_compare_T}, it can be seen that there are
only a few prediction values with deviations, while other
predictions are very accurate.

\FloatBarrier

\bibliography{reference}{}
\bibliographystyle{aasjournal}




\end{document}